\documentclass[12pt]{article}
\begin{document}
\newcommand{\be}{\begin{equation}}
\newcommand{\ee}{\end{equation}}
\newcommand{\ba}{\begin{eqnarray}}
\newcommand{\ea}{\end{eqnarray}}
\renewcommand{\thefootnote}{\arabic{footnote}}
\newenvironment{technical}{\begin{quotation}\small}{\end{quotation}}

\sloppy

\begin{center}
\centering{\LARGE \bf Mechanochemistry\,: an hypothesis for shallow earthquakes}
\end{center}
\vskip 0.5cm
\begin{center}
\centering{\bf Didier Sornette}
\end{center}
\begin{center}
\centering{\bf Department of Earth and Space Science\\ and Institute of Geophysics
and Planetary Physics\\ University of California, Los Angeles, California 90095\\
and\\
Laboratoire de Physique de la Mati\`ere Condens\'ee\\ 
CNRS and Universit\'e des
Sciences, B.P. 70, Parc Valrose\\ 06108 Nice Cedex 2, France\\
}
\end{center}

\begin{center}
\centering{\today}
\end{center}

\vskip 1cm
{\bf Abstract\,:}
We advance the novel hypothesis that water in the presence of 
finite localized strain  within fault gouges may lead to the phase 
transformation of stable minerals
into metastable polymorphs of higher free energy density.
Under increasing strain, the transformed minerals eventually become unstable,
as shown from an application of Landau theory of structural phase transitions.
We propose that this instability leads to an explosive
transformation, creating a slightly supersonic shock wave propagating 
along the altered fault core leaving a wake of shaking fragments.
As long as the resulting high-frequency acoustic waves remain of sufficient amplitude to
lead to a fluidization of the fault core, 
the fault is unlocked and free to slip under the effect of 
the tectonic stress, thus releasing the elastic part of the stored energy. 
We briefly discuss observations that could be understood within this
theory.

\vskip 2cm 
in EARTHQUAKE THERMODYNAMICS AND PHASE TRANSFORMATIONS IN THE EARTH'S INTERIOR
- Roman Teisseyre and Eugeniusz Majewski (eds), Cambridge University Press.

\pagebreak
\tableofcontents

\pagebreak

\section{Introduction}

An earthquake is a sudden rupture in the earth's crust or mantle caused
by tectonic stress. This premise is usually elaborated by
models that attempt to account for seismological and geological data as well as
constraints from laboratory experiments. Notwithstanding almost a century
of research since the standard rebound theory of earthquakes was formulated [{\it
Reid}, 1910], the complex nature and many facets of earthquake 
phenomenology still escape
our full understanding and is reflected in the disappointing progresses on
earthquake prediction [{\it Keilis-Borok}, 1990; {\it Mogi}, 1995; 
{\it Kanamori}, 1996; {\it Geller et al.},
1997]. 

We first review three important paradoxes, the strain paradox, the stress
paradox and the heat flow paradox, 
that are difficult to account for in the present stage of understanding 
of the earthquake processes, either individually or when taken together. 
Resolutions of these
paradoxes usually call for additional assumptions on the nature of
the rupture process (such as novel modes of deformations and ruptures)
prior to and/or during an earthquake, 
on the nature of the fault and on the effect of trapped
fluids within the crust at seismogenic depths.

We then review the evidence for the
essential importance of water and its interaction with the modes of deformations
[{\it Kirby}, 1984; {\it Hickman et al.}, 1995;
{\it Evans and Chester}, 1995; {\it Thurber et al.}, 1997].
We see then present our scenario.

\section{Strain, stress and heat flow paradoxes}

\subsection{The strain paradox}

Within the elastic rebound theory, the order of
magnitude of the horizontal width over which a shear strain develops progressively
across a fault prior to an earthquake is of the order the thickness $\approx
15~km$ of the seismogenic zone. This estimate is robust with respect to the 
many ways with which one can refine the model [{\it Turcotte and Schubert}, 1982;
{\it Sornette}, 1998b].

Modern geodetic measurements are sufficiently precise to test for the existence
of strain localization. The surprise is that there is no geodetic evidence of the
existence of a concentration of shear deformation at the scale of $1-30~km$ 
around major active faults in many situations, even if others exhibit it [{\it Pearson et al.}, 
1995]. The geodetic displacement profiles obtained across section of
faults that have not ruptured in the last decades give an essentially uniform
strain over distances of $150 km$ or more from the fault [{\it Walcott et al.}, 1978;
1979; {\it Shen et al.}, 1996; {\it Snay et al.}, 1996]. 

On the other hand, strain concentrations are observed very
neatly {\it after} a large earthquake [{\it Shen et al.}, 1994\,; {\it Massonnet et
al.}, 1994; 1996]. The observed postseismic relaxation, with concentration in the
vicinity of the fault, is attributed either to viscoelastic relaxation in the
volume of the crust and upper mantle or to afterslip or continued slip on the fault
rupture planes.

\subsection{The stress paradox \label{friction}}

\subsubsection{Statement of the paradox}

There is a large body of literature
documenting the maximum shear stress necessary to initiate sliding as a function of normal
stress for a variety of rock types. The best linear fit defines a maximum
coefficient of static friction $f_s = 0.85 $ [{\it Byerlee}, 1977], with a range
maybe between $0.6$ to $1$ approximately \footnote{Heuze [1983] reviews a large body
of the literature until that time on the various properties of granitic rocks.}. The
range has been confirmed by stress inversion of small faults slip data in situ with
result $0.6 \pm 0.4$ [{\it Sibson}, 1994]. 

There is a big discrepancy between the stress threshold of $200 ~MPa$ at $10 ~km$ depth
implied by this value of friction measured for rocks in the
laboratory and the stress that is available in nature to trigger an earthquake.
To develop such a large stress, shear strain should reach
 $\epsilon \approx 3 ~10^{-3}$ [{\it Sornette}, 1998b]
 demanding an intense localization over an horizontal
 width of a few kilometers. As already discussed, this is not usually observed.

In addition, the stress of the San
Andreas fault zone deduced from a variety of techniques is found to be low and
close to perpendicular to the fault [{\it Zoback et al.},
1987]. This is in contradiction with the frictional sliding model. A more detailed
tensorial analysis assuming hydrostatic pore pressure gives the depth-averaged
shear strength of faults in the brittle continental crust under a typical 
continental geotherm and strain rate of about $35~MPa$ in normal faulting,
$150~MPa$ in thrust faulting and $60 ~MPa$ in strike-slip faulting [{\it Hickman}, 1991].
A set of stress-field indicators, including 
borehole breakouts, earthquake focal mechanisms
and hydraulic fractures, suggests that many (but not all) active faults are
sliding in response to very low levels of shear stress [{\it Zoback}, 1992a; 1992b].
Also noteworthy is
the paradox of large overthrusts [{\it Brune et al.}, 1993], namely that thrust 
faults exhibit an orientation too close to the horizontal to obey the usual 
friction law, thus also requiring an anomalously small resistance to friction.
Such low-angle faulting is observed in many places and is in contradiction with Anderson's
classification of faults [{\it Anderson}, 1951] and with the friction 
theory usually used for the other modes of
faulting. The conclusion on this weakness of faults is contrary to 
traditional views of fault strength based upon laboratory experiments and 
creates a serious problem as one cannot rely directly on the knowledge 
accumulated in the laboratory. 

The situation is made even more confusing when one refers to the occasional observations
of very high stress drops ($30~MPa$ to $200~MPa$) in moderate earthquakes [{\it Kanamori}, 
1994]. This would indicate that the stress drop (and therefore the absolute stress level)
can be very large over a scale of a few kilometers. But then we should see large strain
and anomalous heat flux due to frictional heating (see below).

\subsubsection{Proposed resolutions}

To account for these puzzles, many suggestions have been made. Let us mention in
particular that Chester [1995]
has proposed a multi-mechanism friction model including cataclitic flow, localized
slip and solution transfer assisted friction in order to describe the mechanical
behavior of the transitional regime at midcrustal depths. 
Blanpied et al. [1995] also use the multi-mechanism friction model for frictional slip
of granite. They stress that extrapolating the laboratory results to conditions not
encompassed by the data set (i.e. to approach the conditions in the crust) is
uncontrolled as many mechanisms are competing in a complex way. 
The solution which is often proposed to resolve this problem is to invoke fluid pore
pressure. The presence of a fluid decreases the normal stress
and thus the shear stress necessary to reach the threshold while the most favorable
slipping direction is unchanged. This last condition ensures the compatibility with
the in-situ stress inversion measurements.
The problem however remains to find a mechanism for pressurizing fault fluids
above the hydrostatic value towards the lithostatic value in short time scales
compatible with earthquake cycles [{\it Lachenbruch}, 1980; {\it Rice}, 1992;
{\it Blanpied et al.}, 1992; 1995; {\it Scholz}, 1990;
{\it Brace}, 1980; {\it Lockner and Byerlee}, 1993].
To summarize, it seems that invoking fluid overpressure to weaken the effective
friction coefficient serves a single purpose and creates novel difficulties to
interpret other observations. We refer to [{\it Sornette}, 1998b] and the references
therein for more information.

\subsection{The heat flux paradox}

\subsubsection{Statement of the paradox}

The heat flow paradox [{\it Hickman}, 1991] in a seismically active
region was first proposed by Bullard [1954]\,: to allow for large earthquakes, a
fault should  have a large friction coefficient so that it can store a large amount of
elastic energy and overpass large barriers. However, if the dynamical friction
coefficient is large, large earthquakes should generate a large quantity of heat
not easily dissipated in a relatively insulating earth. Under repetition of
earthquakes, the heat should accumulate and either result in localized melting
(which should inhibit the occurrence of further earthquakes) or develop a high
heat flow at the surface. 0bservations over the entire state of California
have shown the absence of anomalous heat flow across the major faults [{\it Henyey
and Wasserburg}, 1971; {\it Lachenbruch and Sass}, 1980; 1988; 1992; 1995;
{\it Sass et al.}, 1992].

\subsubsection{Proposed resolutions}

A standard explanation is that no significant heat is generated because the
dynamical friction coefficient is low on the most rapidly slipping faults in
California. Two classes of models are then usually proposed. In the first class, 
the low friction is produced dynamically during the event itself. Various mechanisms
are invoked, such as crack-opening modes of slip [{\it Brune et al.}, 1993; {\it
Anooshehpoor and Brune}, 1994; {\it Schallamach}, 1971], dynamical collision effects 
[{\it Lomnitz-Adler}, 1991; {\it Mora and Place}, 1994; {\it Pisarenko and Mora},
1994; {\it Schmittbuhl et al.}, 1994], self-organization of gouge particles under
large slip [{\it Lockner and Byerlee}, 1993; {\it Scott et al.}, 1994; {\it Scott},
1996], acoustic liquifaction [{\it Melosh},
1996] and so on. In the second class of models, the fault has a low friction before
the onset of the event. This may be due to the presence of low-strength clay
minerals such as Montmorillonite (the weakest of the clay minerals) [{\it Morrow
et al.}, 1992], to an organized gouge structure similar to space filling bearings
with compatible kinematic rotations [{\it Herrmann et al.}, 1990], to the presence
of phyllosilicates in well-oriented layers [{\it Wintsch et al.}, 1995] or
to the existence of a hierarchical gouge and fault structure
leading to renormalized friction [{\it Schmittbuhl et al.}, 1996].
Much attention
has also been devoted to the role of overpressurized fluid [{\it Lachenbruch}, 1980;
{\it Byerlee}, 1990; {\it Rice}, 1992; {\it Blanpied et al.}, 1992; 1995; {\it Sleep
and Blanpied}, 1992; {\it Moore et al.}, 1996] close to the lithostatic pressure.
The problem is that fluid pressure close to lithostatic value 
implies fluid trapping and absence of
connectivity with larger reservoirs and the upper surface. Scholz
[1992b] has also noticed that the fluid pressure scenario has one major problem that
remains to be resolved, namely that a brittle material can never resist a pore
pressure in excess of the least compressive stress $\sigma_3$ without drainage
occurring by hydrofracturing. 
Other suggestions to explain the absence of an anomalous heat flux near active faults
include geometric complexity in the San Andreas fault at depth, hydrothermal circulation
and missing energy sinks (see [{\it Hickman}, 1991] and references therein).

\section{Chemistry\,: mineral alteration and chemical transformation}

\subsection{Alteration}

The first element of our approach consists in recognizing
that the preparation for an earthquake
starts at all scales, down to the molecular level. Under the action of a slowly increasing
tectonic strain and in the presence of water which tends to concentrate within fault zone
discontinuities [{\it O'Neil and Hanks}, 1980; 
{\it Thurber et al.}, 1997]
and within defects in the minerals, rocks undergo a
progressive hydration and lattice distorsion. Chemical reactions are
slowly taking place within the rock fabrics. These processes
involve the interaction of water molecules, both intact and decomposed into hydroxyl
and hydrogen components by the exchange of electrons with the
$SiO_4^{4-}$ building block of silicates and also with the
impurities within the rock minerals. For instance, phyllosilicates can precipitate in
rock systems in the presence of water, provided magnesium is present [{\it Wintsch et
al.}, 1995]. Crystal
plastic deformation (through dislocation movement) is enhanced by an atomic scale
interaction of a component of the water with the $Si-O$ bond structure.
As a consequence, the minerals deform progressively, storing an increasing
density of dislocations that are nucleated and stabilized by the
presence of hydroxyl and other impurities in their cores. Some
deformation is displacive, {\it i.e.} corresponds to a distortion of the
crystalline mesh as for instance in the local $Si-O-Si$ unit flipping induced by
the presence of a single water molecule [{\it Jones et al.}, 1992]. Other
deformation is plastic and is due to
the irreversible creation and motion of dislocations. These processes lead to a
weakening of the rock materials. For small strains, this is nothing but hydrolytic weakening.
We are interested in its extension in
the large strain regime. For large strains, the effect of water is
still largely unquantified and work on this problem is a priority in the search
for a deeper understanding of earthquakes and more generally shock metamorphism [{\it
Nicolaysen and Ferguson}, 1990].

This localized deformation and weakening of the rocks is
preferentially concentrated within discontinuities between rock fabrics since it is
the domain where the largest amount of water is thought to be available for
chemical and hydrolytic attack. Shear can then localize
within a few ten-centimeter-thick ultracataclasite zone at the center of fault
cores. These ultracataclasite zones are composed primarily of multiply reworked vein
materials, testifying to repeating rupture and healing [{\it Scholz}, 1992].
In fact, preexisting faults are not necessary, only the preexisting
heterogeneity in the mineral structures that can occur at many scales. The faults
will appear as a consequence of the rupture on these discontinuities. 
Material discontinuities, textures, microcracks, faults are
pervading the earth crust on all scales. It is important to recognize that physical
fields (stress, strain, temperature, fluid pressure...) are, as a result, also
highly heterogeneous.  The formation of fault structures can occur
by repeated earthquakes [{\it Sornette et al.}, 1994].
Earthquakes are localized on narrow zone in which the
pressure, temperature and water content is particularly favorable for the
preparatory stage. It is possible that faults are also partially selected and
transformed by other processes than those occurring during earthquakes, for
instance by ductile, creep and plastic localization [{\it Od\'e}, 1960; {\it
Orowan}, 1960].

Due to the weakening process and the plastic deformation and softening of the
minerals, a large deformation may be progressively concentrated in a narrow domain
while the intact parts of the rocks deform elastically at a smaller rate.
This alteration does not need to occur in a constant steady state as it is
controlled by the amount of available fluid which may be intermittent in time.
Ultimately, alteration is coupled to the water content through a feedback loop
involving its transport within the crust. If the localized deformation occurring
at depth is coupled mechanically to the surface, it can be seen if an edifice or an
object is sitting right across the discontinuity at the surface (see for instance
[{\it Bolt}, 1993], pages 92-93)\,: the phases of observed continuous slips can be
interpreted as the response of the upper gouge material to reactions in
deeper rocks upon the introduction of water). One observes on some segments of the
San Andreas fault important deformations concentrated in narrow meter-wide zones.
On other segments or on other faults comprising about $90~\%$ of all cases, such
creep is not observed, probably due to the progressive decoupling (dampening) of the
deformations by the overlayers.

The total tectonic strain is not solely accomodated by a given localized zone.
In addition to this localized deformation, large scale strain develops over
the entire loaded region due to tectonic motion. This large scale strain is the
signature at large scale of the tectonic load on the elastic plates. The stress is
transferred to the regions of local alteration and concentrated deformation
through the agency of the large elastic domains which are thus necessarily
deformed. It is thus important to understand that both styles of deformation coexist
as a consequence of stress conservation. Only a fraction of the total
deformation is accomodated in the core of the faults.

\subsection{Transformation of mechanical energy into chemical energy during
alteration via polymorphic phase transformations} 

Under the action of tectonic motion driven by thermal mantle convection, the crust is
storing energy progressively. It also dissipates a part of the driving energy under the
form of ductile and plastic deformations. If balance between input and output is
positive as when the modes of ductile deformation are sub-dominent, more and more
energy is being piled up in the crust. This cannot go on for ever and must be released
suddenly in the form of an earthquake. In the broadest sense, an earthquake is the
sudden release of a fraction of the stored energy, whatever the nature of this
storage. In a nutshell, the earthquake cycle thus
comprises the process of a slow energy storage ending in a brutal energy release. To
understand what causes an earthquake,  we must consider all possible significant forms
of energy storage because the sudden event must release all the forms of energies that
are not dissipated continuously. The form of the energy storage must in turn control
the properties of the earthquake.

In principle, the possible form of energy storage are
\begin{enumerate}
\item elastic energy stored in the
rocks as a result of their elastic deformation under the tectonic loading\,: this is the
standard form considered in most models of earthquakes, as for instance in the elastic
rebound model [{\it Reid}, 1910].
\item electric, dielectric or more generally electromagnetic energy\,: this form of
energy storage is expected to be a very small fraction of the total stored energy due
to the smallness of piezoelectric and electrokinetic conversion factors in the crust.
\item ``chemical'' energy\,: we have seen that rocks are
subjected to pressure, temperature, fluid and chemical conditions that put them in
deformation regimes that are often much more complex than simple elasticity,
especially close to rock discontinuities where fluid can have a very important role.
In the literature, non-elastic modes of deformation are usually taken into account
as factors that control the dynamics rather than the nature of
end products. For instance, the nucleation and start-up of the earthquake rupture are
often described in terms of non-elastic models of deformations.
We propose here that non-elastic modes of deformations may play an
important role also in determining the nature of an earthquake and not only its
dynamics. We thus suggest that chemical energy storage can be significant. Minerals in
fault zones can be considered as analogs of piezo-chemical plants, which convert, in
the presence of water and chemical environment, a part of the elastic energy into
chemical energy that is stored in the form of new compounds.
\end{enumerate}

The proposed scenerio introduces the action of stress and strain in the stability of
minerals. The proposed mechanism belongs to the class of processes known as
{\it mechanochemistry}. Indeed, the motion of a dislocation by that of kink (leading to
plastic deformation) is akin to a local chemical reaction in which an embedded
``molecule'' is dissociated, and then one of the product atoms joins with an atom from
another dissociation to form a new ``molecule'' [{\it Gilman}, 1993]. Now, chemical
reactions can be triggered by mechanical forces in solid phases, because unlike gases
and fluids, solids support shear strain. Shear changes the symmetry of a molecule or a
solid and is thus effective in stimulating reactions, much more so than isotropic
compression [{\it Gilman}, 1996]. The reason for the strained minerals to be able
to transform into metastable minerals lies in the kinetics. For application to the
crust, we need to better understand the constraints in the parameter spaces
(pressure, temperature, water affinity, impurities, strain, strain rate, rock
composition) that may control the chemical transformations.

Let us conclude this part by generalizing.
There is growing recognition that mineral structures can form at much
milder pressures and temperatures than their pure phase diagram would suggest,
when in contact with water or in the presence of anisotropic strain and stress.
For instance, diamonds can now be formed under relatively
low pressure ($100 ~MPa$) and temperature ($500 ~^{\circ}C$) under hydrothermal conditions
[{\it Zhao et al.}, 1997; {\it DeVries}, 1997], while the direct transformation route
from carbon requires a pressure above $12000 ~MPa$ and a temperature of about $2000
~^{\circ}C$. Another case in point is that the application of
uniaxial stress along prefered directions in quartz minerals results in the
appearance of a new crystalline phase, where all silicon atoms are in fivefold
coordination [{\it Badro et al.}, 1996]. The stress threshold for the transition
is lowered in this case by the application of the anisotropic stress. Novel
behavior can also appear such as the instability of
the melt-crystal interface [{\it Grinfeld}, 1986; {\it Thiel et al.}, 1992]
in the presence of nonhydrostatic strain. These recent discoveries are
suggestive of the wealth of new phenomena that are possible when chemistry and/or
phase transformations are coupled to anisotropic mechanical deformations.

\section{Dynamics\,: explosive release of chemical energy}

\subsection{Energetics}

The metastable minerals are forming under
tectonic stress and thus grow with mineralogic orientations governed by
those of the applied stress. They also start to deform until they
become unstable and convert back to more stable minerals. 
We propose that the back-reaction may become explosive, due 
the fast release of energy when the minerals become unstable due to the storage of
dislocations and other defects. Only under these circumstances can an explosion occur.
If the minerals are not severely deformed, 
the reversion will be smooth as observed in some experiments
[{\it Green}, 1972]. The experimental challenge, to test our ideas,
is thus to produce sufficient deformation
{\it after} the nucleation of the metastable minerals 
to reach the metastable-unstable threshold at which a small disturbance
explosively modifies its structure. A necessary condition is that the metastable
phase should not recrystallize back spontaneously and progressively, as in 
coesite [{\it Green}, 1972].

Kuznetsov [1966] has determined the thermodynamic conditions for a transition from
a metastable state of matter to a stable equilibrium state to occur as a 
detonation shock. He shows that many first-order (including polymorphic)
phase transitions can lead in
principle to a detonation. The general condition involves the position of the 
adiabat with respect to the isotherm in the pressure-volume diagram of the stable
substance. Barton et al. [1971] and Hodder [1984] has taken up the idea and 
suggested its possible relevance for deep earthquakes. Randall [1966]
has calculated the resulting seismic radiation from a sudden phase transition.

To understand quantitatively and illustrate this process, 
we use the Landau theory of phase
transitions in minerals [{\it Salje}, 1990; 1992; {\it Heine et al.}, 1990]. Landau
theory is simple and economic, as it uses symmetry constraints to derive the free
energy as an expansion of the order parameters, thus minimizing the need for a
detailed mineralogic description. Notwithstanding this simplicity, when the order
parameters are correctly identified, it reproduces experimental observations of phase
transitions in minerals with sufficient accuracy to be useful for many applications.
It can also be extended to account for the dynamics of phase transitions
[{\it Salje}, 1990; 1992].
This formalism should apply to any phase transition from one mineral form to
another as occurs generally for instance in feldspars and other crystalline structures
found in the crust. To keep the discussion as simple as possible, we use the simplest
conceptual model of a coupling between a structural order parameter and the applied
strain. In general, minerals and their structural phase transitions involve several
coupled order parameters $Q_i$, which describe the atomic displacements within the
lattice structure. These parameters form a tensor that couples to the strain tensor 
$\epsilon$.

Consider a structural
transition described by the following free energy\,: 
\be
G = -{1 \over 2} a Q^2 + {1 \over 4} b Q^4 + h Q + d Q \epsilon + {1 \over 2} g
\epsilon^2~~. \label{freener}
\ee
We consider the simplest case of a single scalar order parameter $Q$. More complex
situations do not modify the mechanism and generalization of our discussion to more
complicated situations is straightforward conceptually. The single order parameter $Q$
represents for instance a pure dilational coupling or a pure shear coupling.
${1 \over 2} g \epsilon^2$ is the elastic energy density of the material with
elastic modulus $g$. The coefficient $d$ quantifies the 
strength of the coupling between order parameter and strain. The parameters $a$ and $b$
are phenomenological coefficients for the phase transition. The ``field'' $h$ controls the
breaking of symmetry between the two phases. 

In the absence of strain, the critical
transition occurs at $a=0$ and separates a phase with $Q=0$ (for $a<0$) from a phase
with non-zero $Q$ (for $a>0$) possessing the $Q \to -Q$ symmetry. For $a>0$, $h=0$ and
vanishing strain $\epsilon = 0$, the two phases $Q_{\pm}=\pm \sqrt{a \over b}$ have
the same free energy. The symmetry between $Q$ and $-Q$ is broken by the field $h$.
Keeping $a > 0$ and varying $h$ allows one to describe a first-order transition
between the two phases $Q_+$ and $Q_-$ with a jump in order parameter when $h$ goes
through zero. Note that there is no need of a field if we introduce a $Q^3$ term in
the free energy expansion (\ref{freener}) which breaks the $Q \to -Q$ symmetry. Other
forms up to $Q^6$ have been considered to describe tricritical mineral phase
transitions [{\it Salje}, 1990; 1992; {\it Heine et al.}, 1990]. Here, we use this
simple expression as this is enough to demonstrate the effect. 

We now examine the influence of the
coupling between the order parameter $q$ and the strain field $\epsilon$. 
In fact, strain is recognized as an essential
ingredient in structural phase transitions because elastic strain coupling is the
dominant interaction between atoms and mineral cells in structural phase transitions
[{\it Marais et al.}, 1991; {\it Salje}, 1991; {\it Bratkovsky et al.}, 1995].
For the sake of illustration, we make the stable undeformed mineral 
correspond to $Q_-$ and the metastable phase to $Q_+$, which are obtained
as the two minima of $G$. Note that the presence of strain $\epsilon > 0$ has, 
in this model,
the same effect as an increase of the symmetry-breaking field. 
Other forms of coupling do not have such
a direct equivalence but exhibit however the same qualitative properties. In the
presence of a finite strain $\epsilon > 0$, the two phases have their order parameter
which are solution of ${d G \over dQ} = 0$ (extremum of the free energy) with ${d^2
G \over dQ^2} > 0$ (stability condition\,: local minimum). 
The first condition gives the cubic equation 
\be
Q^3 + A Q + B = 0~~,
\label{cubiceq}
\ee
with $A = {a \over b}$ and $B = {h + d \epsilon \over b}$. Three cases occur [{\it
Beyer}, 1991]. 
\begin{enumerate}
\item ${B^2 \over 4} + {A^3 \over 27} < 0$\,: eq.(\ref{cubiceq}) has three real roots,
two of which are stable. This is the regime where the two phases $Q_-$ and $Q_+$ are
locally stable and the phase $Q_+$ with the highest free energy is metastable. This
condition holds for  $\epsilon < \epsilon^* \equiv -{h \over d} + {2 \over \sqrt{27}}
{a^{3 \over 2} \over d}$.
\item ${B^2 \over 4} + {A^3 \over 27} > 0$, i.e. $\epsilon > \epsilon^*$\,:
eq.(\ref{cubiceq}) has one real solution and two conjugate complex solutions. Only the
real one has a physical meaning and corresponds to the unique stable phase.
\item ${B^2 \over 4} + {A^3 \over 27} = 0$, i.e. $\epsilon = \epsilon^*$\,:
eq.(\ref{cubiceq}) has three real roots of which at least two are equal. For this
critical value $\epsilon^*$ of the strain, the metastable state $Q_+$ becomes
unstable as ${d^2 G \over dQ^2}|_{Q_+}$ vanishes and transforms into the stable phase
$Q_-$. 
\end{enumerate}
Figure 1 schematically represents the whole process. Starting from a double well
configuration with $Q_-$ more stable than $Q_+$, the deformation applied to the $Q_-$
phase creates a higher energy state which eventually becomes
comparable to $Q_+$. As a consequence, the system transforms into the
metastable $Q_+$. As the strain continues to increase, the free energy landscape
deforms until a point where $Q_+$ becomes unstable and the mineral transform back into
{\it undeformed} $Q_-$. 

The energy released in the chemical transformation is estimated as follows.
The free energy difference between the metastable phase and the
stable mineral is taken of the typical order of
 $10^2~kJ/kg \approx 2 \times 10^8~J/m^3$. Consider a volume
of $40~km$ by $10~km$ by $0.1~m$, corresponding to the core of a fault activated
by an earthquake similar to Loma Prieta (1989 earthquake, $M_W = 6.9$)
[{\it Kanamori and Satake}, 1990] or Landers earthquake [{\it Cohee and Beroza} 1994], in which
alteration has transformed about $1\%$ of the minerals into the energetic metastable
phase. This volume can then release $10^{14}~J$ from the explosive phase
transformation (compared to
the energy released, say, by the Hiroshima bomb equal to $13~kT$ of TNT $\approx
5.2 \times 10^{13}~J$).  For Loma Prieta, the seismic
radiation energy is in the range [{\it Kanamori et al.}, 1993; {\it Houston}, 1990;
{\it Choy and Boatwright},
1995] $E_S = 10^{15-16}~J$. Another estimate is given by the 
static elastic calculation using a simple shear rupture model [{\it Knopoff}, 1958] giving
a total released energy $E_T \approx  {\pi \over 8} G d^2 L$. This calculation 
assumes that the stress drop is equal to the average stress level prior to the event.
For $d \approx 1.6~ m$ over a rupture length $L \approx 40 ~km$, we get 
$E_T \approx 1.3 \times 10^{15}$.

\subsection{Explosive shock propagation}

The first chemical
transformation is a slow process as it is fed by a slowly increasing
deformation. In contrast, the transformation 
from the deformed metastable
phase into underformed minerals is not slow and has a dynamics which is expected to be
linear in time (i.e. a front velocity can be defined). The reason is the following.
If a structural phase transition occurs at non-constant chemical
composition, the dynamics is in general diffusive and thus slow. However, if the
composition is constant (as is expected here for structural phase transitions
in minerals), the transformation may occur as a front propagation (so-called ``massive''
transformation) [{\it Christian}, 1965]. Among massive transformations, the
displacive transformations can even occur in volume at an extremely fast rate
since it involves
only local bond rotations (as in quartz $\alpha$ $\to$ quartz $\beta$ for instance). This last
situation does not apply in general to structural phase transformations that require 
rupture of bonds and not only atomic rotations. A ``massive'' transformation is
obtained by the propagation of curved, flexible boundaries which move with variable
speeds and have the ability to cross grain boundaries. The velocity is orders of
magnitude higher than that of a reaction involving long-range diffusion. During
transformation, migration over only a few interatomic distances is required.

There is a more fundamental reason for the fast propagation of the explosive front. 
The point is that the products left in the wake of the front are highly fragmented
minerals and not the well-structured arrangements that would be the result of a slow
diffusive-limited phase transformation. The usual slow velocity of phase transformations
is due to this diffusion-limited nature which is absent here in an explosive process.

If the transformation can be slow or fast depending on the pressure and temperature
conditions, it has been shown that the chemical reaction at the atomic level is not
essentially different even when explosive [{\it Lonsdale}, 1969].
In the mineral transformations proposed here, the dynamics should be even faster than for
standard ``massive'' transformations in alloys. The reason is that no thermal
activation is needed as the free energy barrier is made to vanish by the increase of
strain. At the atomic level, large strains allow a delocalization of electronic
charges helped by impurities which lead to a new mineral structure 
[{\it Gilman}, 1992; 1993; 1995b; 1996]. This is a situation which is very similar to
what happens with  solid explosives. Indeed, explosive substances are nothing but
species storing chemical energy in metastable chemical configurations that is
released suddenly. Here, an energetic solid substance (the metastable phases)
releases energy quickly by transforming to a more stable low energy substance. It
has been proposed [{\it Gilman}, 1995a] that intense deformation by bending of atomic
bonds occur in a very narrow zone of atomic scale that can propagate at velocities
comparable to or even higher than the velocity of sound in the initial material.

Let us show that the propagation of the explosion front occurs at supersonic speed. 
Let us consider an metastable phase that has grown aligned with a coherent
crystalline structure under the influence
of the global stress and strain fields. Imagine the following simple model.
A one-dimensional chain is made of atoms of mass $m$
linked to each other by energetic links of spring constant $k$
which, when stressed beyond a limit, rupture by releasing 
a burst of energy $\Delta g$ converted into
kinetic energy transmitted to the atoms. 
Initially the chain of atoms is immobile. Suppose that the first atom on the left 
is suddenly brought to a position that entails the rupture of the first bond. This rupture
releases the energy $\Delta g$ that is converted into kinetic energies of the
atom fragment that is expelled to the left and of the next atom to the right
which becomes the new left-boundary of the chain. 
All atoms along the chain start to move progressively due to the
transmission of the motion by the springs. Now,
due to the impulsive boost $\sqrt{\Delta g \over m}$ that the boundary 
atom received, it will 
eventually stress the bond linking it to the next atom towards its rupture threshold. When this
occurs, it is expelled by the energy that is released and the next atom forming the new
boundary is itself boosted suddenly by the amount $\sqrt{\Delta g \over m}$. It is then clear
that this leads to a shock propagating at a velocity larger than the sound velocity 
$\sqrt{k \over m}$ since the atoms are receiving boosts that accelerate their motion faster
than what would be the propagation by the springs with the usual acoustic wave velocity.
Taking the continuous limit, 
the resulting supersonic shock velocity $U$ is given by an adaptation of the formula proposed by
Gilman [1995]\,:
\be
U^2 =  c^2 + \Delta G~,
\ee
where $c$ is the longitudinal (P-wave) velocity (around $5000~m/s$), and 
$\Delta G \approx 10^2~kJ/kg$ is a typical value for the free energy release by the
transformation.  This yields $U \approx 1.003~ c$.

 We present in the appendix 1 a calculation of the properties of such a
front propagation during an explosion in a one-dimensional tube configuration. The
calculation highlights the dependence of the front velocity on the energy released by
the explosion. The appendix 1 treats a solid-solid explosion with
a front separating a material with high energy ahead of it from a region of denser and
strengthened material with lower energy behind it. The front propagation is found
supersonic with respect to the wave velocity (here in 1D) in the weaker material ahead
of it but remains subsonic with respect to the wave velocity behind it. 

Associated to the mechanical transformation and the associated elastic waves, we also
expect electrical signals to be generated. Indeed, the metastable phase destabilization
and transformation to other minerals occurs, as already stressed, in coherent aligned
crystals. As a consequence, 
a net non-vanishing piezoelectric effect should appear and the transformation
to other minerals must induce a significant electromagnetic pulse. The elasto-electric
coupled modes in a piezoelectric material are calculated in the appendix 2.
The appendix 3 revisits the calculation of the appendix 1 in the presence of the 
electric coupling process.

\section{Dynamics\,: the genuine rupture}

\subsection{Explosive fluidization and unlocking of the fault}

The explosive transformation of the deformed unstable energetic mineral phase
is violent and leads to fragmentation
with the generation of intense shaking due to high frequency sound waves that
remain trapped in the loosened low acoustic impedence core of the fault.
We argue that this leads to acoustic fluidization [{\it Russo et al.}, 1995].
Melosh [1996] has proposed acoustic fluidization as a mechanism for the low dynamical
friction of faults (the initial unlocking of the fault is not
described in his scenario). We differ from him in that 
the source of the high frequency acoustic modes is not the rupture propagation
but rather the explosive chemical phase transformation. Furthermore,
the acoustic pressure does not need
to reach the overburden pressure to produce a fluidization of the fault. Sornette [1998b]
has shown that there is a problem with Melosh's mechanism because it
predicts a slip velocity during an earthquake more than two orders of magnitudes
smaller than the typical meters per second for observed earthquakes. 

It is possible to save this mechanism by invoking that
the acoustic pressure does not need
to reach the overburden pressure but only a small
fraction $\eta$ of it, in order to liquidify the fault. 
Indeed, it is well-established experimentally 
[{\it Biarez and Hicher}, 1994] that the elastic modulii of granular media under large cyclic
deformations are much lower than their static
values. This effect occurs only for sufficiently large amplitudes of the cyclic
deformation, typically for strains $\epsilon_a$ above $10^{-4}$. 
At $\epsilon_a = 10^{-3}$, the elastic modulii
are halved and at $\epsilon_a = 10^{-2}$, the elastic modulii
are more than five times smaller than their static values. As a consequence, the
strength of the granular medium is decreased in proportion. Extrapolating
these properties to the crust, we need 
to estimate the strain created by the acoustic field. The acoustic
pressure is related to the acoustic particle velocity $v$ by $p = \rho c v$.
Assuming $p = \eta \rho g h$, this yields $v = \eta g h/c \approx 12~m/s$ for
$p \approx 200~MPa$, a density  $\rho = 3~10^3~kg/m^3$, a velocity 
$c=4000~m/s$ and $\eta = 0.1$. At a frequency $f$, 
this corresponds to an acoustic wave displacement $u_a = 
v/2 \pi f \approx 2~10^{-3}~m$ at $f \sim 10^3~Hz$.
 The corresponding strain $u_a/w$ is $\sim 2~10^{-3}$
for a gouge width $w$ of the order of one meter [{\it Melosh}, 1996] over which the
intense shaking occurs. These estimations suggest that Melosh's criterion
that the acoustic stress fluctuations must approach the overburden stress on the fault
for acoustic fluidization to occur
is too drastic and smaller shaking can reduce significantly the fault friction.

The peak deformation strain amplitude $\epsilon_a$ 
of the acoustic waves generated by the
explosion can be estimated from the density change 
${\rho_{metastable} - \rho_{stable} \over \rho_{stable}}$
that we take of the order of $10\%$ during the
the explosive transformation. This leads to
$\epsilon_a \approx 3-4~\%$. The acoustic pressure equals to $\epsilon_a \rho g h$, 
i.e. a few percent of the lithostatic pressure. 
This is almost two order of
magnitude less than the value needed in the acoustic fluidization mechanism proposed
by Melosh [1996].  In the presence of this shaking, the
strength of the granular gouge is drastically decreased 
and the gouge can slip under the applied tectonic stress. In addition, if the
stress-strain curve of the gouge has a maximum at the pressure and temperature conditions of
the seismogenic depth, usually occurring at a deformation of a few percent, the
acoustic waves created by the explosive transformation may even lead to an intrinsic
shear instability, analogous to the localization instability in sand. A third
fluidization mechanism may also add to the instability. When the explosive
transformation occurs, the ensuing shaking starts to deform the gouge material. The
first response to deformation of granular material is to compress,
even if it becomes
dilatant at larger deformations. As a consequence, any interstitial fluid is
first compressed which decreases the friction force between the grains and thus may
lead to a release of the fault, which in turn may start to rupture. This effect is known as the
(standard) fluidization of granular media.

\subsection{Rupture propagation and seismic radiation}

There are two possible
contributions to seismic radiation. First, the phase transformation by
itself radiates seismic
waves\,: Randall [1966] and Knopoff and Randall [1970]
have shown that a sudden change of shear modulus of the
material in the presence of a pre-existing shear strain leads to
 a ``double-couple source'' with the
correct and usual characteristic radiation pattern found in earthquakes at low frequencies.
In fact, it is impossible to discriminate between this source and that of the displacement
dislocation from their first motions. The size of the corresponding
double-couple moment is [{\it Knopoff and Randall}, 1970]
\be
M = 2 \delta \mu ~\delta \epsilon~ V~,
\ee
where $\delta \mu$ is the change in shear modulus, $\delta \epsilon$ is the change
in strain inside the volume $V$ in which the phase transformation has occurred. 
As an order of magnitude, we take a 
change in modulus of the order of $10^{10}~Pa$ and an upperbound
for the change in strain $\delta \epsilon$ one-third the relative change 
in density ($10~\%$).  The volume $V$ is the surface
$S$ of the fault times the width $w \approx 0.1~meter$ 
of its core over which the phase transformation 
occurs. We thus see that this corresponds to the seismic moment of an earthquake
with average slip $w~\delta \epsilon$ equal to a fraction of centimeter, irrespective of the 
size of the fault. This contribution is thus negligible for large earthquakes having
slips of meters or more, but may become significant for small earthquakes which are
also well-recorded. 
In addition to this double-couple component, the change in
bulk modulus and in density radiates isotropically with radial directions 
for the first $P$ motions and no first $S$ motions. However, this is again a small effect.

The second contribution to seismic radiation comes from the mechanical fault slip.
The shaking of the fault core induced by the explosive
 transformation unlocks it; indeed, the fluidization implies that the fault 
 can no longer support the initial loading stress, since its strength tends to 
vanish. This explosive transformation triggers
the fault slip and the fault starts to slip 
under the action of the pre-existing shear stress and radiates
seismic waves. A crucial point in this model is that
the fault slip is not triggered by reaching
a stress threshold (corresponding either to friction unlocking or rupture nucleation),
but rather by a chemical instability\,: as a consequence, 
any level of stress will activate the fault slip when the explosive phase
transformation occurs.
The explosion could also be triggered from another earthquake.
This possibility needs further investigation.

The rupture dynamics is controlled 
by the usual elasto-dynamic equations. 
The rupture propagation lasts as long as the 
high-frequency waves that are trapped within the shaking gouge
with low-acoustic impedence [{\it Harris and Day}, 1997; {\it Li et al.}, 1994]
remain of sufficient amplitude to unlock the fault. The 
detrapping of these waves control the
healing of the fault, and therefore the static stress drop. This mechanism
is expected to produce large variations of static stress drops, depending on
the degree of chemical alteration and storage of chemical energy necessary
to obtain the unlocking of the fault. This is controlled by the detailed
mineralogy in the fault core and the availability of fluids.
The total size of the rupture is 
determined by the extension of the domain over which
the supersonic shock has propagated, which is itself controlled by the 
alteration processes that
have matured the material and stored a suitable amount of chemical energy.
A fault region that is weakly or not altered plays the role of an energy sink
for the explosive shock propagation and will tend to stop it.

\section{Consequences and predictions}

The proposed theory suggests to explain a number of observations into
a coherent framework. 

$\bullet$ {\it Strain, stress and heat flow paradoxes}\,:
There is no need for elastic strain concentration over a scale of about $10~km$ 
(which, as we have reviewed, is usually not observed) and
very localized plastic-ductile strains are expected. There is no need for large
stress to unlock the fault and the low friction is generated dynamically, preventing
heat generation and providing a solution to the heat flow paradox [{\it Lachenbruch
and Sass}, 1980].

$\bullet$ {\it The longer the recurrence time, the larger the stress drop}\,:
Several studies have shown recently a remarkable relationship between the average
slip rate on faults and the stress drop associated to earthquakes occurring on these
faults (see [{\it Kanamori}, 1994] for a review). 
Earthquakes on faults with long repeat times (thousands of years) 
 radiate more energy per unit fault length and have a significant 
 larger dynamical stress drop than those with short repeat times (a few decades to
 a few centuries). In our framework, a longer period gives more time to 
 saturate the chemical energy storage and leads to more ``energetic'' earthquakes.
 
$\bullet$ {\it Seismic P-wave precursors}\,:
Seismic P-wave ``nucleation phases'' have been reported [{\it Beroza and Ellworth}, 1996]
that seem to precede the arrival of the first P-wave
(longitudinal compressive acoustic wave).  These observations are still controversial
[{\it Mori and Kanamori}, 1996], not only because the reported
signals are weak and the effect is hard to establish, 
but also because their presence is essentially ruled out within
the standard pictures. 
The proposed explosive mechanism and resulting shock wave propagating at slightly 
supersonic velocity ($1.003~c$ according to our estimate)
is a natural candidate to rationalize the observation of these seismic P-wave precursors,
if they exist. We predict an advance of about $6$ milliseconds of the precursor to the
first P-wave motion for a propagation over $10~km$ between source and a seismic station.
This seems of the correct order compared to observations 
[{\it Beroza and Ellworth}, 1996]. However, 
we predict a delay proportional to the distance to the station while
 Beroza and Ellworth [1996] find a delay proportional to the fault rupture length.
The finding that the duration of a precursor scales with the size of the earthquake
is in agreement with our model in which
the mechanical rupture occurs on the length of the fault over which the explosive shock has 
occurred and thus we should expect the precursor
to be proportional to the total size of the rupture.

$\bullet$ {\it Tilt anomalies} have sometimes been reported before earthquakes
(for instance before the Haicheng earthquake [{\it Scholz}, 1977; {\it Kanamori}, 1996]). 
The magnitude of the
tilt is usually very small ($1-20~\mu rad$, say) and not always present. The slow phase
transformation 
during the alteration process and chemical energy storage leads to a density change that
produces a weak surface deformation whose direction and amplitude depend on the 
mechanical heterogeneity of the earth. This may suggest a source for the tilt that is 
sometimes observed.
For a phase transformation where $10\%$ of the minerals undergo a
 relative density change of $10~\%$ occurring within the 
fault core over a depth of $10~km$ and a width of $1$ meter, 
the expected tilt anomaly is $\approx 3~\mu rad$ over a distance of $10~km$.

$\bullet$ {\it Earth tides}\,:
The long-term tectonic loading stress rate of the order of $10^{-3}~bar/hr$ is much less
than the stress rate up to $0.15~bar/hr$ 
due to earth tides from gravitational interactions with the Moon and Sun.
Tidal triggering of earthquakes would thus be expected if rupture began soon
after the achievement of a critical stress level. 
The most careful statistical studies have found no evidence of triggering
[{\it Heaton}, 1982; {\it Rydelek et al.}, 1992; {\it Tsuruoka et al.}, 1995;
{\it Vidale et al.}, 1998]. Two existing theories can explain these observations
by invoking high stress rates just before failure. Dieterich's model of state-
and rate-dependent friction predicts high stressing rates accross earthquake nucleation
zones [{\it Dieterich}, 1992]. Alternatively, changes in fluid plumbing of the fault 
system could conceivably be more rapid than tidal strains and may trigger failure
[{\it Sibson}, 1973]. Our theory, which does not attribute the
triggering of an earthquake to a critical stress threshold, is fully compatible
with these observations.

$\bullet$ {\it Seismicity remotely triggered at long distances}\,:
There is strong evidence that the Landers earthquake, June 28, 1992, southern California,
 triggered seismicity at distances equal to many times the source size [{\it Hill et al.},
1993], with a rate which was maximum immediately after passage of the exciting seismic waves.
The problem is that the dynamical stress created by the seismic waves is very small
at these long distances, of the order of the effect of lunar tides ($0.01~MPa$ and less) which
have not been found to be correlated with earthquakes. Sturtevant et al. [1996] have
recently proposed a model in which the earthquakes are triggered by a rapid increase of
pore pressure due to rectified diffusion of small pre-existing gas bubbles in faults
embedded in hydrothermal systems. This model depends on the confluence of several 
favorable conditions, in particular supersaturated gas in water, large mode conversion
occurring in the geothermal field and a very low ($50~m/s$) shear velocity in the 
porous media filled with the bubbly liquid. Alternatively, the chemical instability
we propose should be much more sensitive to high frequency waves than
to low frequency modulations (think of 
the jerky motions that one strives to avoid when manipulating nitroglycerin explosives!).

$\bullet$ {\it Pre-seismic chemical anomalies}\,: Various anomalous precursory chemical 
emissions have been reported recurrently, but not systematically as for
all other proposed precursory phenomena. Nevertheless, it seems to be a phenomenon
worthy of study. For instance, Sato et al. [1986] reported anomalous hydrogen
concentration change in some active faults, and in particular in association with
the Coalinga earthquakes. Tsunogai and  Wakita [1995] reported anomalous 
ion concentrations of groundwater issuing from
deep wells located near the epicenter of the recent earthquake of
magnitude $6.9$ near Kobe, Japan, on January 17, 1995. Similar anomalies
have also been measured for Radon emission [{\it Igarashi et al.}, 1995] for the 
same earthquake. Johansen et al. [1996] presented a thorough statistical analysis
of these precursors for the Kobe earthquakes and concluded that 
these time-dependent anomalies are well fitted by
log-periodic modulations around a leading power law [{\it Sornette}, 1998a], 
qualifying a critical
cooperative behavior. The source of these chemical anomalies may be different but
seem to point to a chemical source, that we tentatively associate with the
chemical transformations that may accelerate close to the instability.

$\bullet$ {\it Electric effects}\,: The metastable crystals are expected to form
with a prefered coherent orientation. For those minerals that do not present 
centers of symmetry, they should thus exhibit a net piezoelectric effect.
Their explosive transformation may then lead to significant co-seismic electric signals.
Precursory electric signals might be associated with partial precursory conversions of the
metastable minerals.
This might provide a scenario for rationalizing precursory observations [{\it Debate on VAN}, 
1996] and help in improving their investigation. 

$\bullet$ {\it Deep earthquakes}\,:
Finally, our theory suggests that deep earthquake, which have been 
proposed to be due to unstable olivine-spinel transformations [{\it Green and Houston},
1995], are not so
different from superficial earthquakes, not only in the nature of their seismic 
radiation but also in their source mechanism. This may call for a re-examination of 
the phase transition mechanism for deep earthquakes
in the light of the action of water and other impurities in presence of finite
strain. 

$\bullet$ {\it Inversion of metamorphic data}\,:
The usual inference of past tectonic conditions from 
the examination of minerals and their corresponding equilibrium
thermodynamic phase diagrams may be questioned in view of the evidence summarized 
here on out-of-equilibrium processes in the presence of water and finite strain.
This may call for a reexamination of the 
models of crustal motions based on inverting metamorphic patterns in fault zones.

\vskip 1cm

{\bf Acknowledgments}: I have benefitted from helpful discussions with Y. Brechet,
J.M. Christie, P.M. Davis, P. Evesque, J. Gilman, M. Harrison, H. Houston, D.D. Jackson, W. D.
Ortlepp, G. Ouillon, E. Riggs, T. Tsullis and J. Vidale. 
Useful correspondences with J.-C. Doukhan, J. Ferguson, G. Martin, J.P. Poirier
are acknowledged. I thank M. Harrison, H. Houston and J. Vidale
for a critical reading of a first version of this manuscript.
L. Knopoff and A. Sornette deserve a special mention for inspiration. Of course, 
all errors remain mine's. I dedicate this work to Jaufray without whom these ideas
would not have come to earth. This is publication 4908 of the Institute of 
Geophysics and Planetary Physics.

\section{APPENDIX 1\,: Explosive shock neglecting electric effects}

Our treatment follows the analysis of Courant and Friedrichs [1985] of shock waves
and we adapt it to the case of solid phase transformations.
We restrict the treatment to a one-dimensional system.
In this simple representation, it does not make difference in the
mathematical description whether we consider a compressive, extensive or shear
(antiplane deformation).
The formalism below ressembles the treatment of Courant and Friedrichs [1985] for
wave propagation of finite amplitude waves in elastic-plastic materials, but differ
in one essential point, namely in [{\it Courant and Friedrichs}, 1985]
the stress-strain characteristics of plastic
material is of the weakening type and thus does not allow for shocks. We 
study the opposite case where a shock occurs.

We consider a long bar parallel to the $0x$ axis. The bar is initially deformed
uniformly with a strain $s^T$ which is the sum of an elastic part $s_1$ and
a ductile-plastic part $s_1^p$\,: $s^T = s_1 + s_1^p$. In the first initial solid
phase, the elastic modulus is $G_1$. A perturbation is brought at one
extremity of the bar and a new crystalline phase is nucleated and the phase
transformation propagates along the bar. The new solid phase has a
different {\it larger} elastic strain $s_2$ and thus smaller plastic strain $s_2^p
= s^T - s_2$, since the total strain remains fixed. The structural transformation
can thus be viewed as a transmutation of plastic into elastic strain by the local
atom rearrangements within the crystalline mesh. The new phase has a different
{\it stronger} elastic modulus $G_2$. This transformation is thermodynamically
favorable because the total energy, sum of the internal energy and of the elastic
energy, decreases even if the elastic energy increases. The plastic part is
taken into account by the internal thermodynamic energy which determines what is the
stable solid phase. 

Our goal is to calculate the
characteristics of the propagation of the phase transformation which will turn out
to be a shock. For the sake of being specific, we consider an extensive strain (the
conclusions are identical for compressive or shear strains). 
We will be concerned only with the
elastic part of the strain as it is the only one which contributes to the elastic
energy. This is done to simplify the treatment which is presented as a plausibility
demonstration. This assumption amounts to neglect the variation of the fraction of
the density accomodated by the plastic part of the deformation.
All our presentation below thus substracts the plastic
deformation.  

If $a$ is the initial position of an
atom in the bar, it becomes $x(a,t)$ under some elastic deformation. The
elastic component of the strain is given by  
\be
s = {\partial x \over \partial a} - 1 ~~~~~.
\label{strain}
\ee
If $\rho_0$ is the initial density, conservation of mass reads $\rho_0 da = \rho
dx$ and thus 
\be
{\rho_0 \over \rho} = {\partial x \over \partial a} = 1 + s~~~~.
\label{relation}
\ee

The velocity of an atom along the bar is 
\be
u = {\partial x \over \partial t}~~~~~.
\label{velocu}
\ee

The equation of motion is 
\be
\rho {\partial u \over \partial t} = {\partial \sigma \over \partial x}~~~~.
\label{ettgvv}
\ee
The r.h.s. of (\ref{ettgvv}) can be written 
${\partial \sigma \over \partial x} = {\partial \sigma \over \partial a}
{\partial a \over \partial x} = {\partial \sigma \over \partial s}
{\partial s \over \partial a} {\partial a \over \partial x}$. It makes sense to
consider the derivative of the stress with respect to strain, as we consider only
the elastic strain and impose fixed plastic deformations in the two solid phases.
Using the relation ${\rho_0 \over \rho}  = 1 + s$ and (\ref{strain}), we get the
wave equation
\be
{\partial^2 x \over \partial t^2} = g^2 {\partial^2 x \over \partial a^2} ~~~~~,
\label{eropo}
\ee
with
\be
g \equiv \sqrt{{1 \over \rho_0} {\partial \sigma \over \partial s}}~~~.
\label{azsqxw}
\ee
Note that $g$, called the rate of change or shift rate, is usually different from
the sound velocity $v \equiv \sqrt{-{\partial \sigma \over \partial \rho}}$. It is
straightforward to check that
\be
\rho_0 g = \rho v~~~~,
\label{azqsxc}
\ee
which implies that $g$ increases when the material is denser. In the elastic range
where ${\partial \sigma \over \partial s} = G$ is constant, we get $g = \sqrt{G
\over \rho_0}$ which is constant and $v= {\rho_0 \over \rho} g$ which varies with
the density $\rho$ and thus with the deformation. Note that in the limit of small
deformations, these differences can be neglected and $g$ and $v$ become indentical.
Here we keep the distinction as it is important for finite deformations as occurs
in ``massive'' structural transition in which the atoms move over a distance equal to a
finite fraction of the lattice mesh.

The structural solid-solid transition is modelled in this framework through the
form of the function $\sigma(s)$. Before the transformation, we assume a
relatively weak elastic solid $\sigma = G_1 s$. In the neighborhood of a deformation
threshold $s^*$, the modulus crosses over to a larger one $G_2$ and the
characteristics becomes  $\sigma = G_2 s$ for $s > s^*$. 
 This schematic dependence summarizes the nature
of the solid-solid transition between a weak solid with elastic strain $s_1$ to a
stronger solid with larger strain $s_2$.

Consider a bar of material deformed uniformly with strain $s_0$ everywhere along
$0x$. Suppose that a localized perturbation or inhomogeneity produces a local
deformation larger than $s^*$ at the left boundary of the bar. This perturbation
is taken to represent the local nucleation of the stronger solid
phase. The question we address is that of the growth of this new phase.
Qualitatively, the density perturbation will start to advance to the right in the
lighter phase. Since the phase rate $g(s)$ increases with $s$, the largest
deformations propagate the fastest. An initial smooth
disturbance will progressively steepen and a shock will eventually form. The
essential condition for the formation of a shock is thus the increase of $g(s)$
with $s$. We imagine the extremity of the bar to be suddenly extended and to move
with constant velocity $u_P$ in the immobile crystal ahead. No matter how small
$u_P$ is, the resulting motion cannot be continuous because a continuous motion
would imply a forward-facing simple wave, {\it i.e.} a centered simple wave, in
order to achieve a discontinuous change of velocity at the origin. However, the
material velocity through a centered simple wave becomes negative if it vanishes
ahead of the simple wave. Therefore, no adjustment to the positive piston velocity
is possible by continous motion. The answer to this problem is that a shock front
appears, moving away from the extremity of the bar with a constant supersonic speed
(with respect to $v_1 = \sqrt{G_1 \over \rho_0}$), uniquely determined by the
density and the velocity of the quiet crystal and the piston speed. We note
that the shock condition is not obeyed in the analysis of Courant and Friedrichs
[1985] of wave propagation of finite amplitude waves in elastic-plastic materials. 
The shock studied here is indeed due to the rigidifying condition which is absent
in elastic-plastic materials. 

In the presence of a shock discontinuity, we have to write the conservation
equations in integral form as the usual differential formulation is not adapted to
treat the discontinuity at the shock. We note $a_1(t)$ (resp. $a_2(t)$) a point on
the left (resp. right) side of the shock.

The conservation of mass reads 
\be
{d \over dt} \int_{a_1(t)}^{a_2(t)} \rho dx  =  0~~~~.
\ee
 The conservation of momentum reads
\be
{d \over dt} \int_{a_1(t)}^{a_2(t)} \rho u dx  =  p(a_1, t) - p(a_2,t)
 = \sigma(a_2, t) - \sigma(a_1, t)~~~~.
\ee
$p$ is the external imposed pressure, which is opposed in sign by the internal
stress. The conservation of energy reads
\be
{d \over dt} \int_{a_1(t)}^{a_2(t)} \rho (e + {1 \over 2} u^2 + {1 \over 2}
{\sigma s \over \rho}) dx  = p(a_1, t) u(a_1, t) - p(a_2,t) u(a_2, t)~~~~, 
\ee
where $e$ is the internal energy of the crystal.
The r.h.s. corresponds to the work of the external pressure at the extremities of
the bar. 

If all fields are continuous, we retrieve the usual equations of motion. Here, we
assume, from the preceeding considerations, that a point of discontinuity exists
within the bar at position $x = \xi(t)$ and we note the velocity of the front 
\be
{d \xi \over dt} = U(t)~~~~~.
\label{shockvelo}
\ee
Consider an integral of the type 
$J =  \int_{a_1(t)}^{a_2(t)} \Psi(x,t) dx$, where $\Psi$ is discontinuous at
$x=\xi$ . Then, as shown in [{\it Courant and Friedrichs}, 1985], ${d J \over dt}$
must be calculated as 
\be
{d J \over dt} = {d \over dt} \int_{a_1(t)}^{\xi(t)}
\Psi(x,t) dx +  {d \over dt} \int_{\xi(t)}^{a_2(t)} \Psi(x,t) dx~,
\ee
 with keeping in
mind that  $\Psi$ is discontinuous at $x=\xi$. In the limit where $a_1 \to
a_2$ while still keeping the condition $a_1 < \xi < a_2$, we get the fundamental
shock equation
\be
{d J \over dt} = \Psi_2 V_2 - \Psi_1 V_2~~~~~,
\label{fundejfdh}
\ee
where $\Psi_{1(2)} \equiv \Psi(a_{1(2)}, t)$ and 
\be
V_{1(2)} \equiv u_{1(2)} - U~~~~~.
\label{defvited}
\ee

Applying (\ref{fundejfdh}) to the above conservation equations leads to the usual
fundamental shock equations\,:
\be
\rho_1 V_1 = \rho_2 V_2 \equiv m~~~~,
\label{rfcxv}
\ee
where $m$ is the mass flux across the shock.
\be
m u_1 + p_1 = m u _2 + p_2  \equiv c~~~~~,
\label{jjfkhfkj}
\ee
where $c$ is the impulse. The equation of energy conservation yields, after some
manipulations,
\be
e_1 + {1 \over 2} V_1^2 + {p_1 \over \rho_1} + {1 \over 2} {\sigma_1 s_1 \over
\rho_1} = e_2 + {1 \over 2} V_2^2 + {p_2 \over \rho_2} + {1 \over 2} {\sigma_2 s_2
\over \rho_2}~~~~.
\label{dfxcdh}
\ee
Using (\ref{jjfkhfkj}) and (\ref{rfcxv}), we find that
\be
m^2 = - {p_1 - p_2 \over {1 \over \rho_1} - {1 \over \rho_2}}~~~~,
\label{cvxk}
\ee
which reduces to the wave velocity in the limit of very weak shocks. We thus
recover the fact that a sound wave can be interpreted as an infinitely weak shock.
This must in fact be obvious since the conservation equations then recover the
wave equation directly.

Using $p =-\sigma$, the relation (\ref{relation}) between the strain $s$ and
density $\rho$ and the two linear elastic Hooke's law ($\sigma_1 = G_1 s$ and
$\sigma_2 = G_2 s$) in the two phases on each side of the shock, we rewrite the
energy equation (\ref{dfxcdh}) as 
\be 
e(\rho_2) + {G_2 \over 2\rho_2} ({\rho_0^2 \over \rho_2^2} - 3 
{\rho_0 \over \rho_2} + 2) = 
e(\rho_1) + {G_1 \over 2\rho_1} ({\rho_0^2 \over \rho_1^2} - 3 
{\rho_0 \over \rho_1} + 2)~~~~.
\label{dfxyhcdh}
\ee
The two different solid phase structures of the crystal give us $e(\rho_1)$ and
$e(\rho_2)$ (or their difference). We can usually also determine the elastic
coefficients $G_1$ and $G_2$ of the two mineral phases. The equation 
(\ref{dfxyhcdh}) thus determines $\rho_2$ as a function of $\rho_1$. From this, we
deduce the flux $m$ from (\ref{cvxk}). And with (\ref{rfcxv}), we get $V_1$ and
$V_2$ and then deduce $u_2$ (assuming $u_1=0$ corresponding to a material at rest
ahead of the shock) from (\ref{jjfkhfkj}). We then deduce the shock velocity $U =
-V_1 = u_2 - V_2$.

Under extension and with our conditions for the shock that $0 < s_1 < s_2$ and
$\sigma_1 < \sigma_2$, we verify that 
$m^2$ given by (\ref{cvxk}) is positive. Indeed, $m^2$ can be written $m^2 = \rho_0
{\sigma_1 - \sigma_2 \over s_1 - s_2} > 0$ for compression or extension as the
sign results from the strengthening character of the transition, {\it i.e.}
${\partial \sigma \over \partial s} > 0$. Note that a usual ductile rheology has
the opposite sign and therefore cannot develop a shock 
[{\it Courant and Friedrichs}, 1985].

Using (\ref{defvited}) together with (\ref{rfcxv}), and with the initial repose
condition that $u_1 = 0$, we get the formula for the shock velocity 
\be
U = {|m| \over \rho_0} = \sqrt{{1 \over \rho_0}{\sigma_2 - \sigma_1 \over s_2 -
s_1}}  ~~~~~, 
\label{nnnnm}
\ee
where we take the absolute value as $m$ is negative, 
since the flux of matter goes from the right to left. We can compare the shock
velocity $U$ with the sound wave velocities $v_1$ and $v_2$ given by 
$v_{1(2)} \equiv \sqrt{-{\partial \sigma_{1(2)} \over \partial \rho_{1(2)}}} = 
{\rho_0 \over \rho_{1(2)}} \sqrt{G_{1(2)} \over \rho_0}$. It is easier to compare 
$\rho_0 U^2 = {G_2 s_2 - G_1 s_1 \over s_2 - s_1}$ with $\rho_0 v_{1(2)}^2 =
G_{1(2)} (s_{1(2)}+1)^2$. We construct the difference 
$$
\rho_0 U^2 - \rho_0 v_{1(2)}^2 = 
$$
$${(-1)^{2(1)} \over s_2 - s_1} \biggl(
  s_{2(1)} [G_{2(1)}  - G_{1(2)} (s_{1(2)}+1)^2] + 
G_{1(2)}(s_{1(2)}^3 + 2 s_{1(2)}^2 \biggl)~~~~~.
$$
We verify that $\rho_0 U^2 - \rho_0 v_1^2 >0$ for a very large set of
parameters while $\rho_0 U^2 - \rho_0 v_2^2 < 0$ is realized if ${G_2 \over G_1}$
and/or ${s_2 \over s_1}$ are sufficiently large. This condition is realized for
instance with $G_2 = 2 G_1$ and $s_2 = 3 s_1 = 0.3$ while if $s_2 = 2 s_1 = 0.2$,
it is not. The shock regime corresponds to the situation where 
$v_1 < U < v_2$.

\section{APPENDIX 2\,: Elastic-electric coupled wave}

For the simplicity of the exposition, we restrict to a 1D model and express the
elastic and piezoelectric equation in scalar form. This does not preclude from the
fact that piezoelectricity occurs in materials lacking a center of symmetry.
Generalization to the full tensorial expressions involves straightforward
manipulations. The elastic material obeys Hooke's law relating the stress $\sigma$
to the strain $S$\,: \be \sigma = G S  ~~~~, \label{straing}
\ee
where $G$ is the elastic modulus. The electric analog relates the electric
induction $D$ to the electric field $E$\,:
\be
D = \epsilon E~~~~,
\label{electds}
\ee
where $\epsilon$ is the dielectric coefficient of the material. In the presence of 
a non-vanishing piezoelectric effect, (\ref{straing}) and (\ref{electds}) become
coupled through the piezoelectric equations\,:
\be
S = G^{-1} \sigma + d E~~~~~,
\label{one}
\ee
\be
D = d \sigma + \epsilon E~~~~~,
\label{two}
\ee
where $d$ is the piezoelectric coupling coefficient ($=2~~10^{-12} ~CN^{-1}$ for
quartz).

In order to derive the propagative modes, we supplement these two constitutive
equations (\ref{one},\ref{two}) with the fundamental elastic and electric
equations.
\be
\rho {\partial^2 u \over \partial t^2} = {\partial \sigma \over \partial x}~~~,
\label{eqgt}
\ee
where $u$ is the displacement at position $x$.
Together with (\ref{straing}) valid in the absence of piezoelectric coupling and
using $S = {\partial u \over \partial x}$, we get the standard wave equation 
\be
{1 \over v^2} {\partial^2 u \over \partial t^2} = 
{\partial^2 u \over \partial x^2}~~~~,
\label{waved}
\ee
where the acoustic wave velocity is
\be
v \equiv \sqrt{G \over \rho}~~~.
\label{psioid}
\ee

The two Maxwell equations are
\be
\vec{rot} \vec{E} = -{\partial \vec{B} \over \partial t}~~~~,
\label{maxwella}
\ee
\be
{1 \over \mu} \vec{rot} \vec{B} = {\partial \vec{D} \over \partial t}~~~~~,
\label{maxwellb}
\ee
where $\mu$ is the magnetic permittivity of the material. Together with
(\ref{electds}) valid in the absence of piezoelectric coupling, the elimination of
$\vec{B}$ yields the wave equation 
\be
{1 \over c^2} {\partial^2 \vec{E} \over \partial t^2} = \Delta \vec{E}~~~, 
\label{mawjhh}
\ee
where the electromagnetic wave velocity is
\be
c \equiv {1 \over \sqrt{\mu \epsilon}}~~~.
\label{trfdg}
\ee
In the presence of the piezoelectric coupling given by (\ref{one},\ref{two}), we
get the two coupled equations (keeping the single non-vanishing electric field
component)
\be
{\partial^2 E \over \partial x^2}  =  {1 \over c^2} (1 - {G \over \epsilon} d^2) 
{\partial^2 E \over \partial t^2} + d G \mu {\partial^3 u \over \partial x \partial
t^2}~,
\label{hhsslql}
\ee
\be
{\partial^2 u \over \partial x^2}  =  {1 \over v^2}  {\partial^2 E \over \partial
t^2} + d {\partial E \over \partial x}~. \label{twoh}
\ee
We recover (\ref{waved}) and (\ref{mawjhh}) in absence of coupling $d=0$. The
strength of the elastic-electric coupling (EEC) wave is measured by the
dimensionless parameter ${G \over \epsilon} d^2$. For quartz, $\epsilon = 4.5
\epsilon_0$ and we take the value $G \approx 30 GPa$. With
$d =2~~10^{-12} CN^{-1}$, we get ${G \over \epsilon} d^2 \approx 3~~10^{-3}$, {\it
i.e.} a small coupling. 

We look for propagative modes $E = E_0 e^{ikx -\omega t}$, $u=u_0 e^{ikx -\omega
t}$. Inserting in (\ref{twoh},\ref{hhsslql}), we obtain the dispersion relation, by
the condition of vanishing the determinant, under the form of a quadratic equation in
$\omega^2$ as a function of $k^2$\,:
\be
(1 - {G \over \epsilon} d^2) \omega^4 - (c^2+v^2) k^2 \omega^2 + c^2v^2 k^4 = 0~~~,
\label{yhbcg}
\ee
whose general solution is
\be
\omega^2_{\pm} = {(c^2+v^2) k^2 \pm \sqrt{\Delta} \over 
2 (1 - {G \over \epsilon} d^2)} ~~~,
\label{geneu}
\ee
where $\Delta = (c^2+v^2) k^4 - 4 c^2 v^2 k^4 (1 - {G \over \epsilon} d^2)$.
Since the coupling ${G \over \epsilon} d^2 << 1$, we can write with a good
approximation $\Delta = k^4 (c^2 - v^2)^2$. Within this approximation, there is a
fast mode propagating at close to the velocity of light and a slow mode
propagation slightly slower than sound. To first order in ${v^2 \over c^2}$, we get
\be
\omega^2_+ = c^2 k^2 {1 + {G \over \epsilon} d^2 {v^2 \over c^2} \over
1 - {G \over \epsilon} d^2} \approx c^2 k^2 (1 + {G \over \epsilon} d^2)~~~~,
\label{rdefcv}
\ee
and
\be
\omega^2_- = v^2 k^2 {1 + 2 {G \over \epsilon} d^2 {v^2 \over c^2} \over
1 - {G \over \epsilon} d^2} \approx v^2 k^2 (1 + {G \over \epsilon} d^2)~~~~.
\label{roijcv}
\ee
Note that $\omega$ is real since we have neglected dissipation.
From (\ref{hhsslql},\ref{twoh}), we get the relationship between the electric
field amplitude and elastic displacement of the two modes\,:
\be
E_O^+  = - i {c^2 \over v^2} {k \over d} u_0^+~~~~, \label{etry}
\ee
\be
E_O^-  =  i {Gdk \over \epsilon} u_0^-~~~~. \label{twfghoh}
\ee
For a similar elastic amplitude $u_0^+ = u_0^-$, we see that ${E_O^- \over E_O^+}
= {G d^2 \over \epsilon} {v^2 \over c^2} << 1$. This shows that the fast mode is
essentially electromagnetic with negligible elastic deformation and the slow mode is
mainly elastic with very weak electromagnetic fields.

An arbitrary perturbation of the displacive type will
decompose onto these two modes which will thus be excited and propagate in the
narrow zone where minerals have been aligned to create a net piezoelectric effect.
In particular, the amplitude of the electromagnetic mode is given by the equation
(\ref{two}), namely $\epsilon E_0 = d \sigma$ (in absence of preexisting electric
field). Using $\epsilon = 4.5 \epsilon_0$ with $\epsilon_0 = 8.85~~10^{-12}
Fm^{-1}$ and $d= 2~~10^{-12} CN^{-1}$, we obtain $E_0 \simeq 0.05 \sigma (Pa)$.
For a stress of $10 MPa$, this yields an electric field of $5~~10^5 V/m$. This
numerical value is of course only an order of magnitude as several uncertainties
control its determination, such as the piezoelectric coupling $d$ which is probably
significantly over-evaluated and the stress amplitude. Nonetheless, this
calculation suggests that significant electric fields can be created locally.

\section{APPENDIX 3\,: Structural shock including electric effects}

It is the nature of the structural transition of polarized atoms within the
crystalline structure that electrons are released and exchanged during the
propagation of the shock. We can even infer that, since the hydrolytic weakening
is fundamentally a redox process, the process of the exchange of electrons is a key
participant in the solid-solid phase transition. The release of electrons acts as
a self-fuelling process by the polarization source and local crystal structure
distorsion it induces.
Associated to the elastic shock wave, the modification of the crystalline
structure induces a polarization wave, since electrons are released and exchanged.

The mathematical treatment of the solid-solid phase transformation in
terms of a shock presented in the appendix 1
can be extended to take into account the
electromagnetic coupling. The equations (\ref{rfcxv}) of mass conservation and
(\ref{jjfkhfkj}) of momentum conservation remain the same. The equation
(\ref{dfxcdh}) of energy conservation is modified into
\be
e_1 + {1 \over 2} V_1^2 + {p_1 \over \rho_1} + {1 \over 2} {\sigma_1 s_1 \over
\rho_1} + {1 \over 2} {D_1^2 \over \epsilon_1}= e_2 + {1 \over 2} V_2^2 + {p_2 \over
\rho_2} + {1 \over 2} {\sigma_2 s_2 \over \rho_2} + {1 \over 2} {D_2^2 \over
\epsilon_2}~~~~,
\label{rtfvbp}
\ee
where $D_{1(2)}$ is the electric induction in phase $1$ (resp. $2$),
and $\epsilon_1$ (resp. $\epsilon_2$) is the dielectric constant in phase $1$ (resp.
$2$). We have neglected the magnetic contribution as the phase transition is
mainly associated with a transfer of electric charges and is thus an electric effect.

The appendix 2 has calculated the coupled elastic-electric modes in a piezoelectric
medium. As an order of magnitude, we can estimate the amplitude of the electric
induction discontinuity at the shock from the piezoelectric equation (\ref{two})
of the appendix 1 with $E=0$ (negligible external electric field)\,: $D = d \sigma$.
Note that the discontinuity of the electric induction reflects the abrupt
change of the polarization between the two phases. This gives the following
equation which replaces (\ref{dfxyhcdh})\,:
\be 
e(\rho_2) + {G_2 \over 2\rho_2} ({\rho_0^2 \over \rho_2^2} - 3 
{\rho_0 \over \rho_2} + 2) + {d^2 \over \epsilon_1} G_1 ({\rho_0 \over \rho_1} -
1) =  e(\rho_1) + {G_1 \over 2\rho_1} ({\rho_0^2 \over \rho_1^2} - 3 
{\rho_0 \over \rho_1} + 2) + {d^2 \over \epsilon_2} G_2 ({\rho_0 \over \rho_2} -
1)~~~~.
\label{dgvbyhcdh}
\ee

The discontinuity in stress and strain at the shock phase transition produces an
impulse both mechanical and electrical. This generates seismic radiations and
electric signals. This provides a natural mechanism for the
generation of electric signals associated to earthquake rupture. The electric
signal leads to a polarization wave which propagates in the crust and can be
detected on the surface in favorable conditions.

One should distinguish between two types of electric signals\,: 1) electromagnetic
signals of very low frequency propagate as electromagnetic waves\,; 2)
polarization currents are carried along conducting paths within the crust. The
first phenomenon is a fast process that occurs essentially with the explosive
transformation while the second one is a slow phenomenon and can be detected after the
conduction has brought the impulsive charge to the detectors.

We now calculate the amplitude of the first electromagnetic process. For this, we
use the wave equation (\ref{hhsslql},\ref{twoh}) of 
the appendix 2, writing it in 3D, neglecting 
the small term ${G \over \epsilon} d^2 << 1$ and replacing the coupling term
$d G \mu {\partial^3 u \over \partial x \partial t^2}$ by  $d \mu {\partial^2
\sigma \over \partial t^2}$ which acts as a source term
corresponding to the passage of the shock, with stress discontinuity 
$\sigma_2 - \sigma_1$. We thus have ${\partial \sigma \over \partial t} =
{\sigma_2 - \sigma_1 \over \Delta t}$, where $\Delta t$ is the duration of the
shock, {\it i.e.} the time it takes for the shock to pass over a point. In the
inviscid limit, the width of the shock shrinks to zero and $\Delta t \to 0$, thus
leading to  ${\partial \sigma \over \partial t} \to
(\sigma_2 - \sigma_1) \delta(t-t^*(\vec{r}))$, where $t^*(\vec{r})$ is the time of
arrival of the shock at $\vec{r}$. We thus get finally the following wave equation with a
source term due to the shock\,:
\be
\Delta E -  {1 \over c^2} {\partial^2 E \over \partial t^2} = 
d \mu (\sigma_2 - \sigma_1) \delta'(t-t^*)~~~~,
\label{onedfdxt}
\ee
where $\Delta$ is the Laplacian and $\delta'$ is the derivative of the Dirac
function. The formal solution of this equation is obtained by the method of Green
function and reads\,:
\be
E(\vec{r}, t) = {\mu d \over 4 \pi} (\sigma_2 - \sigma_1) \int_{S} dS 
{\delta'\biggl( t-{|\vec{r} - \vec{r}'| \over c} - {|\vec{r}'| \over U} \biggl) 
\over |\vec{r} - \vec{r}'|}~~~~.
\label{refvcp}
\ee
We have replaced $t^*$ by ${|\vec{r}'| \over U}$, which is the time of arrival of
the shock at position $\vec{r}'$ on the fault. The integral is carried over the
surface covered by the shock. Since $U < c$ ($U$ is comparable to a sound wave
velocity, while $c$ is the speed of light in the medium),  we can carry out the
integration with the dirac function expressed in the $\vec{r}'$ variable and obtain
in 1D\,:
\be
E(x, t) = {\mu d \over 4 \pi} ~{\sigma_2 - \sigma_1 \over t - {x \over U}}~~~~.
\label{zerfvcp}
\ee
This gives a signal with a long tail whose peak propagates at the shock
velocity. The duration of the electric signal is thus directly proportional to the
length of the rupture, and thus gives a direct information of the size of the
event.

In 3D, we define $0x$ as the direction along the long axis of the rupture, $0y$ is
the direction parallel to the rupture and perpendicular to $0x$ and $0z$ is
perperdicular to the rupture plane. Solving the integral in (\ref{refvcp}) with the
dirac function, we get
\be
E(\vec{r}, t) = {\mu d \over 4 \pi} (\sigma_2 - \sigma_1) \int dy' 
{x-x'(y') \over |\vec{r} - \vec{r'}|^2}~~~~,
\label{refuhfdbcvcp}
\ee
where $x'(y)$ is solution of 
\be
Ut - x' = {U \over c} \sqrt{(x-x')^2 + (y-y')^2 + z^2}~~~,
\ee
and $\vec{r'}$ is expressed at the point of coordinates $(x'(y'), y',0)$.
We notice that $dy' {x-x'(y') \over |\vec{r} - \vec{r'}|^2} =  dx' {dy' \over dx'}
{d |\vec{r} - \vec{r'}|^{-1} \over dx'}$. We thus obtain
\be
E(\vec{r}, t) = {\mu d \over 4 \pi} (\sigma_2 - \sigma_1) {dy' \over
dx'}|_{B} {1 \over |\vec{r} - \vec{r'}_b|} 
- {\mu d \over 4 \pi} (\sigma_2 - \sigma_1) \int dx' {{d^2 y' \over dx'^2} 
 \over |\vec{r} - \vec{r'}|}~,
\label{refuuuuww}
\ee
where the index $B$ refers to the contribution of the boundaries.
$\vec{r}'_b$ denotes the position of the beginning and end of the rupture. There is
thus a specific component of the signal radiated from the two edges of the rupture.

\vskip 1cm
\section{References}
{\small
Anderson, E.M., The dynamics of faulting, 2nd ed., (Oliver and Boyd, Edinburgh, 1951).

Anooshehpoor, A., and Brune J.N., Frictional heat generation and seismic
radiation in a foam rubber model of earthquakes, 
{\it Pure and Applied Geophysics, 142},735-747, 1994. 

Badro, J., J.-L. Barrat and P. Gillet, Numerical simulation of $\alpha$-quartz
under nonhydrostatic compression\,: memory glass and five-coordinated crystalline
phases, {\it Phys. Rev. Lett., 76}, 772-775, 1996.

Baeta, R.D. and K.H.G Ashbee, Slip systems in quartz, I-Experiments;
II-Interpretation, {\it Am. Mineral., 54}, 1551-1582, 1970.

Barton, A.F.M., A.P.W. Hodder and A.T. Wilson, Explosive or detonative phase
transitions on a geological scale, {\it Nature, 234}, 293-294, 1971.

Beroza, G.C., and W.L. Ellworth,  Properties of the seismic nucleation phase, {\it
Tectonophysics, 261}, 209-227, 1996.

Beyer, W. H., editor, {\it CRC standard mathematical tables and formulae},
29th ed.,  Boca Raton : CRC Press, 1991.

Biarez, J., and P.-Y. Hicher, Elementary mechanics of soil behavior\,: saturated
remoulded soils, Rotterdam ; Brookfield, VT : A.A. Balkema, 1994.

Blanpied, M.L., D.A. Lockner and J.D. Byerlee, An earthquake mechanism based on
rapid sealing of faults, {\it Nature, 358}, 574-576, 1992.

Blanpied, M.L., D.A. Lockner and J.D. Byerlee, Frictional slip of granite at
hydrothermal conditions, {\it J. Geophys. Res., 100}, 13045-13064, 1995.

Bolt, B.A., Earthquakes, W.H. Freeman and Co., New York, 1993.

Brace, W.F., Permeability of crustalline and argillaceous rocks\,: status and
problems, {\it International Journal of Rock Mechanics in Mineral Sciences and
Geomechanical Abstracts, 17}, 876-893, 1980.

Bratkovsky, A.M., E.K.H. Salje, S.C. Marais and V. Heine, Strain coupling as the
dominant interaction in structural phase transition, {\it Phase transitions, 55},
79-126, 1995.

Brune, J.N., S. Brown and P.A. Johnson, Rupture mechanism and interface separation
in foam rubber models of earthquakes\,: a possible solution to the heat flow
paradox and the paradox of large overthrusts, {\it Tectonophysics, 218}, 59-67,
1993.

Bullard, E.C., The interior of the earth, pp.57-137 in ''the Earth as a
planet'', G.P. Kuiper, ed., University of Chicago Press, 1954 (see pp. 120-121).

Byerlee,J., Friction of rocks, In Experimental studies of rock friction with
application to earthquake prediction, ed. J.F. Evernden, U.S. Geological Survey,
Menlo Park, Ca, 55-77, 1977.

Byerlee, J., Friction, overpressure and fault normal compression, {\it Geophys. Res.
Lett., 17}, 2109-2112, 1990.

Chopin, C., Coesite and pure pyrope in high-grade blueschists of the Western Alps\,:
a first record and some consequences, {\it Contributions to Mineralogy and Petrology,
86}, 107-118, 1984.

Choy, G.L., and J.L. Boatwright, Global patterns of radiated seismic energy and
apparent stress, {\it J. Geophys. Res., 100}, 18205-18228, 1995.

Christian, J. W., {\it Theory of transformations in metals and alloys\,; an advanced
text book in physical metallurgy}, Oxford, New York, Pergamon Press, 1965.

Cohee, B.P., and G.C. Beroza, Slip distribution of the 1992 Landers earthquake and
its implication for earthquake source mechanisms, {\it Bull. Seismol. Soc. Am., 84}, 
692-712, 1994.

Courant, R., and K.O. Friedrichs, Supersonic flow and shock waves,
Springer-Verlag, New York and Berlin, third edition, 1985.

Darling, R.S., I.M. Chou and R.J. Bodnar, An occurrence of metastable cristobalite in
high-pressure garnet granulite, {\it Science, 276}, 91-93, 1997.

Debate on VAN, Special issue of {\it Geophys. Res. Lett., 23}, 1291-1452, 1996.

DeVries, R.C., Diamonds from warm water, {\it Nature, 385}, 485-485, 1997.

Dieterich, J.H., Earthquake nucleation on faults with rate-dependent
and state- dependent strength, {\it Tectonophysics, 211}, 115-134, 1992. 

Evans, J.P., and F.M. Chester, Fluid-rock interaction in faults of the
San Andreas system - Inferences from San Gabriel fault rock geochemistry
and microstructures, {\it J. Geophys. Res., 100}, 13007-13020, 1995.

Geller, R.J., D.D. Jackson, Y.Y. Kagan and F. Mulargia, Earthquakes cannot be
predicted, {\it Science 275}, 1616-1617, 1997.

Gilman, J.J., Insulator-metal transitions at microindentations, 
{\it J. Mater. Res., 7}, 535-538, 1992.

Gilman, J.J., Shear-induced metallization, 
{\it Philos. Mag. B - Physics of Condensed Matter Structural Electronic
Optical and Magnetic Properties, 67}, 207-214, 1993.

Gilman, J.J., Why Silicon is hard? {\it Science, 261}, 1436-1439, 1993.

Gilman, J.J., Chemical reactions at detonation fronts in solides, 
{\it Philos. Mag. B - Physics of Condensed Matter Structural Electronic
Optical and Magnetic Properties, 71}, 1057-1068, 1995a.

Gilman, J.J., Mechanism of shear-induced metallization, {\it Czechoslovak J.
Physics, 45}, 913-919 1995b.

Gilman, J.J., Mechanochemistry, {\it Science, 274}, 65-65, 1996.

Green II, H.W., Metastable growth of coesite in highly strained quartz, {\it J.
Geophys. Res., 77}, 2478-2482, 1972.

Green II, H.W., D.J. Griggs and J.M. Christie, Syntectonic and annealing recrystallization
of fine-grained quartz aggregate, in {\it Experimental and Natural Rock Deformation}, 
edited by P. Paulitsch, pp. 272-335, Springer, New York, 1970.

Green, H.W., and H. Houston, The mechanics of deep earthquakes,
{\it Ann. Rev. Earth Sci. 23}, 169-213, 1995.

Grinfeld, M.A., Instability of the separation boundary between a
nonhydrostatically stressed elastic body and a melt, {\it Sov. Phys. Dokl. 31},
831-834, 1986.

Harris, R.A., and S.M. Day, Effects of a low-velocity zone on a dynamic rupture,
{\it Bull. Seismol. Soc. Am., 87}, 1267-1280, 1997.

Heaton, T.H., Tidal triggering of earthquakes, {\it Bull. Seism. Soc. Am., 72}, 2181-2200, 1982.

Heine, V., X. Chen, S. Dattagupta, M.T. Dove, A. Evans, A.P. Giddy, S. Marais, S.
Padlewski, E. Salje and F.S. Tautz, Landau theory revisited, {\it Ferroelectrics,
128}, 255-264, 1992.

Henyey, T.L., and G.J. Wasserburg, Heat flow near major strike-slip
faults in California, {\it J. Geophys. Res., 76}, 7924-7946, 1971.

Herrmann, H.J., G. Mantica and D. Bessis, Space-filling bearings,
{\it Phys.Rev.Lett., 65}, 3223-3226, 1990.

Heuze, F.E., High-temperature mechanical, physical and thermal properties of granitic
rocks -- a review, {\it Int. J. Rock Mech. Sci. \& Geomech. Abstr., 20}, 3-10, 1983.

Hickman, S.H., Stress in the lithosphere and the strength of active faults, {\it
Rev. Geophys., 29}, 759-775, 1991.

Hickman, S., R. Sibson and R. Bruhn, Introduction to special section\,: mechanical
involvement of fluids in faulting, {\it J. Geophys. Res., 100}, 12831-12840, 1995.

Hill, D., P.A. Reasenberg, A. Michael, W.J. Arabaz et al., Seismicity remotely
triggered by the magnitude 7.3 Landers, California, earthquake, {\it Science, 260},
1617-1623, 1993. 

Hobbs, B.E., Recrystallization of single crystals of quartz, {\it Tectonophysics,
6/5}, 353-401, 1968.

Houston, H., A comparison of broadband source spectra, seismic energies and 
stress drops of the 1989 Loma-Prieta and 1988 Armenian earthquakes, 
{\it Geophys. Res. Lett., 17}, 1413-1416, 1990.

Igarashi G., Saeki S., Takahata N., Sumikawa K., Tasaka S.,
Sasaki G., Takahashi M. and Sano Y., Ground-water radon anomaly before the Kobe
earthquake in Japan, {\it Science, 269}, 60-61, 1995.

Johansen, A., D., Sornette, H. Wakita, U. Tsunogai, W. Newman and H. Saleur, 
Discrete scaling in earthquake precursory phenomena - Evidence in the Kobe
earthquake, Japan, {\it J. Physique I France, 6}, 1391-1402, 1996.

Jones, R., S. \"Oberg, M.I. Heggie and P. Tole, {\it Ab initio} calculation of the
structure of molecular water in quartz, {\it Philos. Mag. Lett., 66}, 61-66, 1992.

Kanamori, H., Mechanics of earthquakes, {\it Ann. Rev. Earth Planet. Sci. 22}, 207-237,
1994.

Kanamori, H., Initiation process of earthquakes and its implications for
seismic hazard reduction strategy, {\it Proc. Nat. Acad. Sci. USA, 93}, 3726-3731, 1996.

Kanamori, H., and K. Satake, Broadband study of the 1989 Loma-Prieta earthquake, 
{\it Geophys. Res. Lett. 17}, 1179-1182, 1990.

Kanamori, H., J. Mori, E. Hauksson, T.H. Heaton, L.K. Hutton and L.M. Jones,
Determination of earthquake energy release and $M_L$ using Terrascope, {\it Bull.
Seism. Soc. Am., 83}, 330-346, 1993.

Keilis-Borok, V.I., editor, Intermediate-term earthquake prediction\,:
models, algorithms, worldwide tests, {\it Phys. Earth and Planet. Interiors, 61},
1-139, 1990.

Kirby, S.H., Introduction and digest to the special issue on chemical effects of
water on the deformation and strength of rocks, {\it J. Geophys. Res., 89},
3991-3995, 1984.

Knopoff, L., Energy release in earthquakes, {Geophys. J. Roy. Astron. Soc., 1},
44-52, 1958.

Knopoff, L., and M.J. Randall, The compensated linear-vector dipole\,: a possible 
mechanism for deep earthquakes, {\it J. Geophys. Res. 75}, 4957-4963, 1970.

Hodder, A.P.W., Thermodynamic constraints on phase changes as earthquake source
mechanisms in subduction zones, {\it Phys. Earth and Planet. Int., 34}, 221-225, 1984.

Kuznetsov, N.M., Detonation and gas-dynamic discontinuities in phase transitions
of metastable substances, {\it Soviet Physics JEPT, 22}, 1047-1050, 1966.

Lachenbruch, A.H., Frictional heating, fluid pressure and the resistance of
fault motion, {\it J. Geophys. Res., 85}, 6097-6112, 1980.

Lachenbruch, A.H., and J.H. Sass, Heat flow and energetics of the San
Andreas fault zone, {\it J. Geophys. Res., 85}, 6185-6222, 1980.

Lachenbruch, A.H., and J.H. Sass, The stress heat-flow paradox and thermal results 
from Cajon Pass, {\it Geophys. Res. Lett. 15}, 981-984, 1988.

Lachenbruch, A.H., and J.H. Sass, Heat flow from Cajon Pass, fault strength
and tectonic implications, {\it J. Geophys. Res., 97},
4995-5015, 1992.

Lachenbruch, A.H., J.H. Sass, G.D. Clow and R. Weldon, Heat flow at
Cajon Pass, California, revisited, {\it J. Geophys. Res., 100},
2005-2012, 1995.

Li, Y.G., K. Aki, D. Adams, A. Hasemi and others, Seismic waves trapped in the fault
zone of the Landers, California, earthquake of 1992, {\it J. Geophys. Res., 99}, 
11705-11722, 1994; {\it 99}, 17919-17919, 1994.

Linde, A.T., M.T. Gladwin, M.J.S. Johnston, R.L. Gwyther and R.G. Bilham, A slow
earthquake sequence on the San Andreas fault, {\it Nature 383}, 65-68, 1996.

Liu, L.-G., Bulk moduli of $SiO_2$ polymorphs\,: Quartz, coesite and stishovite, 
{\it Mechanics of Materials 14}, 283-290, 1993.

Lockner, D.A., and J.D. Byerlee, How geometrical constraints contribute to the
weakness of mature faults, {\it Nature, 363}, 250-252, 1993.

Lomnitz-Adler, J., Model for steady-state friction, 
{\it J. Geophys. Res., 96}, 6121-6131, 1991.

Lonsdale, K., The geometry of chemical reactions in single crystals, 
in Physics of the solid state, S. Balakrishna, M. Krishnamurthi and B.R. Rao, eds.,
Academic Press, London and New York, 43-61, 1969.

Marais, S., V. Heine, C. Nex and E. Salje, Phenomena due to strain coupling in phase
transitions, {\it Phys. Rev. Lett. 66}, 2480-2483, 1991.

Martin, G., and P. Bellon, {\it Solid State Physics, 50}, 189-331, Ehrenreich and
Spaepen eds, Academic press, 1997.

Massonnet, D., K. Feigl, M. Rossi and F. Adragna, Radar interferometric mapping of
deformation in the year after the Landers earthquake, 
{\it Nature, 369}, 227-230, 1994.

Massonnet, D., W. Tatcher and H. Vadon, Detection of postseismic fault-zone
collapse following the Landers earthquake, {\it Nature, 382}, 612-614, 1996.

Melosh, H.J., Dynamical weakening of faults by acoustic fluidization,
{\it Nature, 379}, 601-606, 1996.

Mogi, K., Earthquake prediction research in Japan, {\it J. Phys. Earth 43}, 533-561, 1995.

Moore, D.E., D.A. Lockner, R. Summers, M. Shengli et al., Strength of
chrysotile-serpentinite gouge under hydrothermal conditions - can it explain
a weak San Andreas fault? {\it Geology, 24}, 1041-1044, 1996.

Mora, P and D. Place, Simulation of the frictional stick-slip
instability, {\it Pure and Applied Geophysics, 143}, 61-87, 1994.

Mori, J., and H. Kanamori, Initial rupture of earthquakes in the 1995 Ridgecrest,
California sequence, {\it Geophys. Res. Lett., 23}, 2437-2440, 1996.

Morrow, C., B. Radney and J. Byerlee, Frictional strength and the effective
pressure law of Montmorillonite and Illite clays, in Fault mechanics and transport
properties of rocks, B. Evans and T.-F. Wong, eds., Academic Press, London, 69-88
1992.

Nicolaysen, L.O., and J. Ferguson, Cryptoexplosion structures, shock
deformation and siderophile concentration related to explosive venting of
fluids associated with alkaline ultramafic magmas, {\it Tectonophysics, 171},
303-335, 1990.

Od\'e, H., Faulting as a velocity discontinuity in plastic deformation, {\it The
Geological Society of America Memoir, 79}, A symposium on Rock Deformation, D.
Griggs and J. Handin, eds., 293-321, 1960.

Okay, A.L., X. Shutong and A.M.C. Seng\"or, Coesite from the Dabie Shan eclogites,
central China, {\it Eur. J. Mineral., 1}, 595-598, 1989.

O'Neil, J.R., and T.C. Hanks, Geochemical evidence for water-rock interaction along
the San-Andreas and Garlock faults in California, {\it J. Geophys. Res. 85}, 6286-6292, 1980.

Orowan, E., Mechanism of seismic faulting,  {\it The
Geological Society of America Memoir, 79}, A symposium on Rock Deformation, D.
Griggs and J. Handin, eds., 293-321, 1960.

Ortlepp, W.D., Note on fault-slip motion inferred from a study of
micro-cataclastic particles from an underground shear rupture, {\it Pure Appl. Geophys., 139},
677-695, 1992.

Ortlepp, W.D., {\it Rock Fracture and rockbursts -- an illustrative study} (S.A.
Inst. Min. and Metall. Johannesburg, 1997).

Pearson, C.F., J. Beava, D.J. Darby, G.H. Blick and R.I. Walcott, Strain
distribution accross the Australian-Pacific plate boundary in the central
South Island, New Zealand, from 1992 GPS and earlier terrestrial observations,
{\it J. Geophys. Res. 100}, 22071-22081, 1995.

Pisarenko, D., and P. Mora, Velocity weakening in a dynamical model of friction,
{\it Pure and Applied Geophysics, 142}, 447-466, 1994. 

Randall, M.J., Seismic radiation from a sudden phase transition, 
{\it J. Geophys. Res., 71}, 5297-5302, 1966.

Reid, H.F., The California earthquake of April 18, 1906. The
 mechanics of the earthquake. Vol.  (II of the report of the
 California state earthquake investigation commission (Carnegie Inst. Wash, Pub 
 n87, vol 2)  192 pp (1910).
 
Rice, J.R., Fault stress states, pore pressure distributions and the weakness of
the San Andreas fault, in Fault mechanics and transport properties in rocks (the
Brace volume), ed. Evans, B., and T.-F. Wong, Academic, London, 475-503, 1992.

Russo, P., R. Chirone, L. Massimilla and S. Russo, The influence of the frequency of
acoustic waves on sound-assisted fluidization of beds of fine particles, {\it Powder
Technology, 82}, 219-230, 1995.
 
Rydelek, P.A., I.S. Sacks and R. Scarpa, On tidal triggering of earthquakes at Campi
Flegrei, Italy, {\it Geophys. J. Int., 109}, 125-137, 1992.

Salje, E. K. H., Phase transitions in ferroelastic and co-elastic crystals\,: an
introduction for mineralogists, material scientists, and physicists,
Cambridge, England; New York : Cambridge University Press, 1990.

Salje, E. K. H., Strain-related transformation twinning in minerals, {\it Neues
Jahrbuch fur Mineralogie-Abhandlungen, 163}, 43-86, 1991.

Salje, E. K. H., Application of Landau theory for the analysis of phase transitions
in minerals, {\it Phys. Rep., 215}, 49-99, 1992.

Sass, J.H., A.H. Lachenbruch, T.H. Moses and P. Morgan, Heat flow from 
a scientific research well at Cajon Pass, California, 
{\it J. Geophys. Res., 97}, 5017-5030, 1992.

Sato, M., A.J. Sutton, K.A. McGee and S. Russel-Robinson, Monitoring of 
hydrogen along the San Andreas and Calaveras faults in Centra California in 1980-1984,
{\it J. Geophys. Res., 91}, 12315-12326, 1986.

Schallamach, A., How does rubber slide? {\it Wear 17}, 301-312, 1971.

Schmittbuhl, J., J.P. Vilotte and S. Roux, Dynamic friction of self-affine
surfaces, {\it J. Phys. II France, 4}, 225-237, 1994.

Schmittbuhl, J., J.P. Vilotte and S. Roux, Velocity weakening friction -- A
renormalization approach, {\it J. Geophys. Res., 101}, 13911-13917, 1996.

Scholz, C.H., A physical interpretation of the Haicheng earthquake prediction, {\it
Nature 267}, 121-124, 1977.

Scholz, C.H., Shear heating and the state of stress on faults, {\it J. Geophys. Res. 85},
6174-6184, 1980.

Scholz, C.H., The mechanics of earthquakes and faulting, Cambridge University
Press, Cambridge, 1990.

Scholz, C.H., Weakness amidst strength, {\it Nature, 359}, 677-678, 1992.

Scott, D., Seismicity and stress rotation in a granular model of 
the brittle crust, {\it Nature, 381}, 592-595, 1996.

Scott, D.R., C.J. Marone and C.G. Sammis, The apparent friction of granular fault
gouge in sheared layers, {\it J. Geophys.Res., 99}, 7231-7246, 1994.

Shen, Z.-K., D.D. Jackson, Y. Feng, M. Cline, M. Kim, P.
Fang and Y. Bock, Postseismic deformation following the Landers earthquake,
California, 28 June 1992, {\it Bull. Seism. Soc. Am., 84}, 780-791, 1994.

Shen, Z.-K., D.D. Jackson and B.X. Ge, Crustal deformation across and beyond the
Los Angeles basin from geodetic measurements, 
{\it J. Geophys.Res., 101}, 27957-27980, 1996.

Sibson, R.H., Interactions between temperature and pore fluid pressure during earthquake
faulting and a mechanism for partial or total stress relief, {\it Nature, 243}, 66-68, 1973.

Sibson, R.H., An assessment of field evidence for `Byerlee' friction, {\it
Pure Appl. Geophys., 142}, 645-662, 1994.

Sleep, N.H., and M.L. Blanpied, Creep, compaction and the weak rheology of major
faults, {\it Nature, 359}, 687-692, 1992.

Smith, D.C., and M.A. Lappin, Coesite in the Straumen kyanite-eclogite pod, Norway,
{\it Terra Research, 1}, 47-56, 1989.

Snay, R.A., M.W. Cline, C.R. Philipp, D.D. Jackson et al., Crustal velocity field
near the big bend of California San Andreas fault,
{\it J. Geophys.Res., 101}, 3173-3185, 1996.

Sornette, D., Discrete scale invariance and complex dimensions, 
{\it Physics Reports, 297}, 239-270,1998a.

Sornette, D., Earthquakes: from chemical alteration to mechanical rupture, 
submitted to Physics Reports, 1998b (http://xxx.lanl.gov/abs/cond-mat/9807305)

Sornette, D., P. Miltenberger and C. Vanneste, 
Statistical physics of fault patterns self-organized by repeated earthquakes, {\it Pure
Appl. Geophys. 142}, 491-527, 1994.

Sturtevant, B., H. Kanamori and E.E. Brodsky, Seismic triggering by
rectified diffusion in geothermal systems, {\it J. Geophys. Res. 101},
25269-25282, 1996.

Thiel, M., A. Willibald, P. Evers, A. Levchenko, P. Leiderer and S. Balibar,
Stress-induced melting and surface instability of $^4He$ crystals, {\it Europhys.
Lett., 20}, 707-713, 1992.

Thurber, C., S. Roecker, W. Ellsworth, Y. Chen, W. Lutter and R. Sessions, 
Two-dimensional seismic image of the San Andreas fault in the northern Gabilan
range, central California\,: evidence for fluids in the fault zone,
{\it Geophys. Res. Lett. 24}, 1591-1594, 1997.

Tsunogai U., and  Wakita H., Precursory chemical changes in ground water - Kobe
 earthquake, Japan, {\it Science, 269}, 61-63, 1995.
(1995).

Tsunogai, U., Wakita, H., Anomalous changes in groundwater chemistry - 
Possible precursors of the 1995 Hyogo-ken Nanbu earthquake, Japan,
{\it J. Phys. Earth, 44}, 381-390, 1996.

Tsuruoka, H., M. Ohtake and H. Sato, Statistical test of the tidal triggering of
earthquakes: contribution of the ocean tide loading effect, {\it Geophys. J. Int., 122},
183-194, 1995.

Tullis, J.A., Prefered orientation in experimentally deformed quartzites, {\it
Ph.D. Thesis, Univ. California, Los Angeles}, 344 pp, 1971.

Tullis, J., J.M. Christie and D.T. Griggs, Microstructures and preferred
orientations of experimentally deformed quartzites, {\it Geol. Soc. Am. Bull.,
84}, 297-314, 1973.

Turcotte, D.L., and G. Schubert, Geodynamics, applications
of continuum physics to geological problems, John Wiley and Sons, New York, 1982.

Vidale, J.E., D.C. Agnew, M.J.S. Johnston and D.H. Oppenheimer, Absence of earthquake
correlation with earth tides; an indication of high preseismic fault stress rate, 
{\it J. Geophys. Res.} in press, 1998.

Walcott R.I. et al., Geodetic strains and large earthquakes in the axial tectonic
belt of North Island, New Zealand, {\it J. Geophys. Res. 83}, 4419-4429, 1978.

Walcott R.I. et al., Strain measurements and tectonics of New Zealand, 
{\it Tectonophysics , 52}, 479, 1979.

Wang, X., and J.G. Liou, Coesite-bearing eclogite from the Dabie mountains in central
China, {\it Geology, 17}, 1085-1088, 1989.

Wintsch, R.P., R. Christoffersen and A.K. Kronenberg, Fluid-rock reaction
weakening of fault zones, {\it J. Geophys. Res., 100},13021-13032, 1995. 

Zhao, D.P., and H. Kanamori, The 1994 Northridge earthquake -- 3-D crustal
structure in the rupture zone and its relation to the aftershock locations and
mechanism, {\it Geophys. Res. Lett., 22}, 763-766, 1995.

Zhao, X.-Z., R. Roy, K.A. Cherlan and A. Badzian, Hydrothermal growth of diamond
in metal$-C-H_2O$ systems, {\it Nature, 385}, 513-515, 1997.

Zoback, M.L., 1st-order and 2nd-order patterns of stress in the lithosphere - The
world stress map project, {\it J. Geophys. Res. 97}, 11703-11728, 1992a.

Zoback, M.L., Stress field constraints on intraplate seismicity in Eastern
North-America, {\it J. Geophys. Res. 97}, 11761-11782, 1992b.

Zoback, M.L., V. Zoback, J. Mount, J. Eaton, J. Healy et al., New evidence on the
state of stress of the San Andreas fault zone, {\it Science, 238}, 1105-1111, 1987.

}

\pagebreak

Figure captions\,:

\vskip 1.5cm

Figure 1: Starting from a double well
configuration with $Q_-$ (undeformed stable mineral) more stable than $Q_+$ 
(metastable mineral), 
the deformation applied to the $Q_-$
phase creates a higher energy state which eventually becomes
comparable to $Q_+$ (this is the effect of inclusion of dislocations, for
instance). As a consequence, the system transforms into the
metastable $Q_+$. As the strain continues to increase, the free energy landscape
deforms until a point where $Q_+$ becomes unstable and the mineral transform back into
{\it undeformed} $Q_-$. 

\vskip 1cm

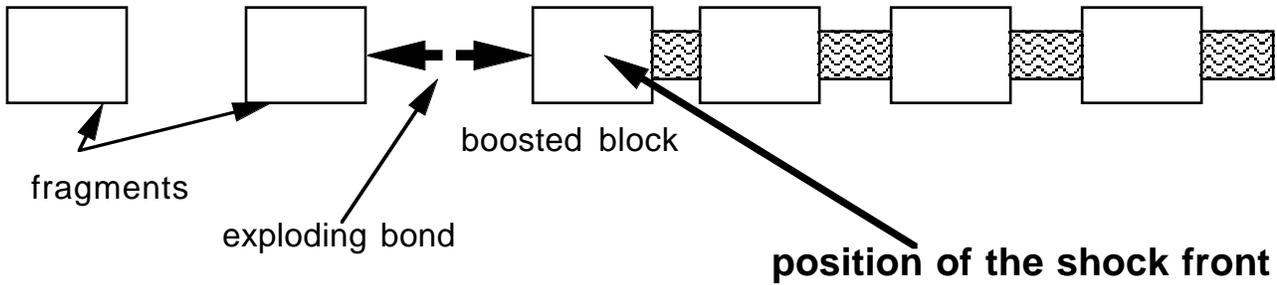
Figure 2: A one-dimensional chain is made of blocks
linked to each other by energetic links
which, when stressed beyond a given deformation threshold,
rupture by releasing a burst of energy converted into
kinetic energy transmitted to the blocks. The figure shows two successive bond
ruptures that lead to velocity boosts to the ejected fragments on the left and to
the boundary blocks.

\end{document}